\documentclass[12pt]{article}
\usepackage{amssymb,amsfonts,amsmath,amsthm}
\usepackage{graphicx}
\usepackage{enumerate}
\usepackage{natbib}
\usepackage{tikz}
\usepackage{bbm}
\usepackage{mathtools}
\usepackage{algorithm,algorithmic}
\usepackage{lscape}
\usepackage{multirow}
\usepackage{etoolbox}
\usepackage{hyperref}
\usepackage{fullpage}

\def\Tp{{\mathcal{T'}}}

\newcommand{\rej}{\text{rej}}

\newcommand{\var}{\text{var}}

\newcommand{\parent}{\text{parent}}

\newcommand{\child}{\text{child}}
\newcommand{\FSP}{{\rm FSP}}
\newcommand{\FSR}{{\rm FSR}}
\newcommand{\FDP}{{\rm FDP}}
\newcommand{\TPP}{{\rm TPP}}
\newcommand{\FDR}{{\rm FDR}}

\newcommand{\depth}{{\rm depth}}
\newcommand{\power}{{\rm power}}
\newcommand{\MSPE}{{\rm MSPE}}

\def\prob{\mathbb{P}}
\def\cL{\mathcal{L}}
\def\E{\mathbb{E}}
\def\ind{\mathbbm{1}}
\def\har{\hbar}
\def\tR{\tilde{R}}
\def\cT{\mathcal{T}}
\def\tgamma{\tilde{\gamma}}
\def\talpha{\tilde{\alpha}}
\def\PTac{\mathcal{P}_{\mathcal{T}_a}^c}
\def\real{{\mathbb R}}

\def\var{\mathrm{Var}}
\def\E{\mathbb{E}}

\def\T{\mathcal T}
\def\L{\mathcal L}
\def\H{\mathcal H}
\def\cov{\mathrm{Cov}}
\def\btheta{\boldsymbol{\theta}}
\def\root{\mathrm{root}}
\def\normal{{\sf N}}
\def\by{\boldsymbol{y}}
 \let\proglang=\textsf \let\code=\texttt
\def\by{\boldsymbol{y}}
\def\bX{\boldsymbol{X}}

\def\beps{\boldsymbol{\varepsilon}}
\def\eps{\varepsilon}
\def\HAT{{\sf HAT}}
\def\bG{\boldsymbol{G}}

\newcommand{\hnull}[1]{\mathcal{H}^0_{#1}}

\newtheorem{propo}{Proposition}[section]
\newtheorem{lemma}[propo]{Lemma}

\newtheorem{proposition}[propo]{Proposition}

\newtheorem{thm}[propo]{Theorem}
\newtheorem{cons}{Constraint}
\newtheorem*{remark}{Remark}

\begin{document}

\bibliographystyle{agsm}

  \title{\bf Controlling the False Split Rate in Tree-Based Aggregation}
  \author{Simeng Shao, Jacob Bien, Adel Javanmard\thanks{
    A.~Javanmard is partially supported by the Sloan Research Fellowship
in mathematics, an Adobe Data Science Faculty Research Award and the
NSF CAREER Award DMS-1844481. J.~Bien was supported in part by NIH
Grant R01GM123993 and NSF CAREER Award DMS-1653017.}\\
Data Sciences and Operations, University of Southern California
}%
  \maketitle

\begin{abstract}
In many domains, data measurements can naturally be associated with the leaves
of a tree, expressing the relationships among these
measurements.  For example, companies belong to industries, which in
turn belong to ever coarser divisions such as sectors; microbes are
commonly arranged in a taxonomic hierarchy from species to kingdoms;
street blocks belong to neighborhoods, which in turn belong to larger-scale regions.
The problem of tree-based aggregation that we consider in this paper asks which of
these tree-defined subgroups of leaves should really be treated as a single
entity and which of these entities should be distinguished from each other.

We introduce the {\em false split rate}, an error measure that
describes the degree to which subgroups have been split when they
should not have been.  We then propose a multiple hypothesis testing
algorithm for tree-based aggregation, which we prove controls this
error measure.  We focus on two main examples of tree-based
aggregation, one which involves aggregating means and the other
which involves aggregating regression coefficients.  We apply this
methodology to aggregate stocks based on their volatility and to aggregate
neighborhoods of New York City based on taxi fares.
\end{abstract}

\noindent%
{\it Keywords:} Multiple testing, false discovery rate, rare features, hierarchy

\section{Introduction}

A common challenge in data modeling is striking the right balance between models that are
sufficiently flexible to adequately describe the phenomenon being
studied and those that are simple enough to be easily interpretable.
We consider this tradeoff within the increasingly common context in which data measurements
can be associated with the leaves of a known tree.  Such data
structures arise in myriad domains from business to science, including
the classification of occupations \citep{soc}, businesses
\citep{naics}, products, geographic areas, and taxonomies in ecology.

Measurements in low-level branches of the tree may share a lot in common,
and so---in the absence of evidence to the contrary---a data modeler
would favor a simpler (literally ``high-level'') description in which distinctions within
low-level branches would not be made; on the other hand, when there is evidence of a difference
between sibling branches, then modeling them as distinct from each
other may be warranted.  We use the term {\em tree-based aggregation} to refer to the process of
deciding which branches' leaves should be treated as the same
(i.e., aggregated) and which should be treated as different from each
other (i.e. split apart).

Tree-based aggregation procedures have been proposed in various
contexts, including regression problems, in which features represent
counts of rare events \citep{Yan2018RareFS} or counts of microbial species
\citep{bien2021tree}, and in graphical modeling \citep{wilms2021tree}.
These approaches focus on prediction and estimation but do not address
the hypothesis testing question of whether a particular split should occur.

We formulate the general tree-based aggregation problem as a multiple
testing problem involving a parameter vector $\btheta^*$
whose elements correspond to leaves of a known tree.  Our goal is to
partition the leaves based on branches of the tree so that the set of parameters in each
group share the same value.  Every non-leaf node has an associated
null hypothesis that states that all of its leaves have the same
parameter value.  Type I errors correspond to splitting up groups
unnecessarily; type II errors correspond to aggregating groups with
different parameter values.

In Section \ref{section:FSR}, we define an error measure, called the
{\em false split rate} (FSR), that corresponds to the fraction of
splits made that were unnecessary.  Within our tree-based setting, we
show that controlling the FSR is  related to controlling the
false discovery rate \citep{BH1995}, with equivalence in the special
case of a binary tree.

In Section \ref{section:Testing}, we propose a tree-based aggregation
procedure that leverages this connection.  Our algorithm proceeds in a
top-down fashion, only testing hypotheses of nodes whose parents were
rejected.  Such an approach to hierarchical testing originates with
\citet{Yekutieli2008}, which lays the foundation for the multiple testing
problem on trees.  Our procedure is closely related to more recent work
by \citet{lynch2016procedures}, which increases power using
carefully chosen node-specific thresholds that depend on where the hypothesis is
located in the hierarchy.  This work was in turn further developed in
\citet{Ramdas2017DAGGERAS}.  Other work involving various forms of a
hierarchy-based multiple testing problem (although not having to do
with aggregation in the sense of this paper) include
\citet{bogomolov2017testing,heller2018post,katsevich2019multilayer}.
While these works focus on FDR control, another line of work uses
hierarchical testing for gradually locating non-zero variables while controlling the family-wise
error rate \citep{Meinshausen,guo2019group}.

In Section \ref{section:two_applications}, we consider two concrete
scenarios where tree-based aggregation is natural.  In the first
scenario, the parameter vector $\btheta^*$ represents the mean of a
scalar signal measured on the leaves of the tree.  In the second
scenario, $\btheta^*$ is a (potentially high-dimensional) vector of
regression coefficients where features are associated with leaves of
the tree.

Finally, we demonstrate through simulation studies (Section
\ref{section:simulation_result}) and real data experiments (Section
\ref{section:real_data}) the empirical merits of our framework and
algorithm.  We consider two applications, corresponding to the two
concrete scenarios of tree-based aggregation.  The first application involves
aggregation of stocks (with respect to the NAICS's sector-industry
tree) based on mean log-volatility.  The second application aggregates
neighborhoods of New York City (with respect to a geographically based
hierarchy) based on a regression vector for predicting taxi drivers'
monthly total fares based on the frequency of different starting locations.

\medskip

\noindent{\bf Notation:}
For an integer $p$, we write $[p] = \{1,2,\dotsc, p\}$. For $a, b\in\real$, we write $a\wedge b$ and 
$a\vee b$ for their minimum and maximum, respectively.
We use $\boldsymbol{e}_i$ to denote the $i$-th standard basis vector. 
For $\boldsymbol{x} \in \mathbb{R}^p$, we define $\left\Vert \boldsymbol{x}\right\Vert_q
= \left( \sum_{j=1}^{p} |x_j|^q \right)^{1/q}$  for $q \geq 0$.  For a
set $S\subseteq [p]$, $\boldsymbol{x}_{S} = (x_i)_{i\in S}$ is the
vector obtained by restricting the vector  $\boldsymbol{x}$ to the
indices in set $S$. We use the term ``tree'' throughout to denote a
rooted directed tree. 
Given a tree $\mathcal
T$ with leaf set $\mathcal L$, we write $\mathcal T_u$ for the subtree rooted at $u\in\mathcal T$
and $\mathcal{L}_u$ for its leaf set.

\section{Problem setup} \label{section:FSR}

\subsection{A multiple hypothesis testing formulation for aggregation} \label{section:tree_parametrization}

Let $\mathcal{T}$ be a known tree with $p$ leaves, each corresponding to a
coordinate of the unobserved parameter vector $\btheta^*\in\real^p$.  We formulate the tree-aggregation task as a multiple hypothesis testing problem: To each internal (non-leaf) node $u$ of the tree we assign a null hypothesis 
\begin{align}\label{eq:HT-tree}
\hnull{u}:\ \text{\rm All elements of~} \btheta^*_{\mathcal{L}_u} \text{\rm ~have the same value},
\end{align}
where $\btheta^*_{\mathcal{L}_u}$ is the subvector of $\btheta^*$
restricted to leaves of the subtree rooted at $u$. Rejecting the null hypothesis $\hnull{u}$ implies that the leaves under $u$ should be further split into smaller groups. Given the way the hypotheses are defined, a logical constraint to impose on the output of a testing procedure is the following:
\begin{cons}\label{cons1}
The parent of a rejected node must itself be rejected.  
\end{cons}
By constraint \ref{cons1}, the set of rejected nodes will then form a subtree
$\mathcal{T}_{\rm rej}$ of $\mathcal{T}$ (sharing the same root as
$\mathcal T$), and furthermore the subtrees rooted at the leaves of $\mathcal{T}_{\rm rej}$ represent the aggregated groups.  
Our goal is to develop testing procedures that result in high quality
splits of the parameters. In order to measure the performance of an aggregation
(or equivalently a set of splits) we propose a new criterion as follows.

\medskip

\begin{figure}
	\centering
	\includegraphics[width=0.7\textwidth]{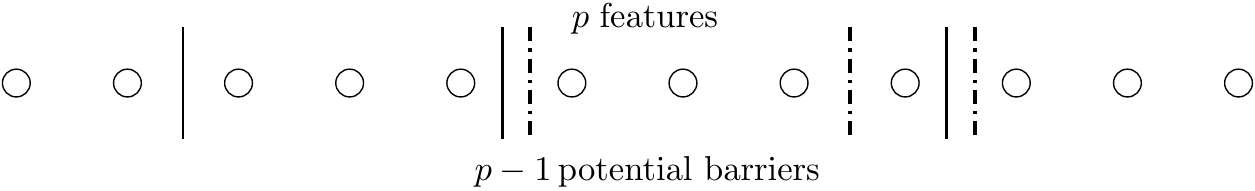}
	\caption{\em\small An example of leaves partition. There are $p=12$ leaves in total, hence $11$ potential barriers. The solid barriers indicate the true splitting of leaves, while the dashed barriers result in the achieved splitting of leaves. In terms of the vector of barriers, $\FDP^b = \frac{1}{3}$ and $\TPP^b = \frac{2}{3}$. In terms of splitting of leaves, $\FSP = \frac{5-4}{4-1} = \frac{1}{3}$ and $\power = 1-\frac{5-4}{4-1} = \frac{2}{3}$. }
	\label{fig:example_FDP}
\end{figure}

\noindent{\bf False Split Rate (FSR).} Recall that we are interested in splits that can be expressed as a combination of branches of the tree $\mathcal{T}$. Therefore if we order the leaves (from left to right), if two leaves are in the same group, then the other leaves between them are also in the same group. For partitioning an ordered sequence of $p$ leaves, we have $p-1$ potential positions for the barriers of groups. We use a vector $\boldsymbol{\vartheta} \in \{0,1\}^{p-1}$ to denote whether the corresponding barrier exists at that position. Each realization of such vector will result in a unique splitting of leaves, and vice versa. Let $\boldsymbol{\vartheta}^*$ and $\widehat{\boldsymbol{\vartheta}}$ respectively denote the corresponding vectors for the true splitting $\mathcal{C}^*$ and an achieved splitting $\widehat{\mathcal{C}}$. 
In Figure~\ref{fig:example_FDP} we give an example of $p=12$ leaves. The solid barriers mark the true splitting, $\boldsymbol{\vartheta}^* = (0,1,0,0,1,0,0,0,1,0,0)$; the dashed barriers mark the achieved splitting, $\widehat{\boldsymbol{\vartheta}} = (0,0,0,0,1,0,0,1,1,0,0)$.

We can view the splitting task as a barrier discovery problem. The
false discovery proportion and true positive proportion can then be written as
\begin{align}\label{eq:FDPb-TPPb}
\FDP^{b} := \frac{|\{j\in[p-1]:\; \vartheta^*_j = 0, \widehat{\vartheta}_j = 1\}|}{|\{j\in[p-1]:\; \widehat{\vartheta}_j = 1\}|}\,,\qquad\TPP^{b} := \frac{|\{j\in[p-1]:\; \vartheta^*_j = 1, \widehat{\vartheta}_j = 1\}|}{|\{j\in[p-1]:\; {\vartheta}^*_j = 1\}|} \,.
\end{align}
Since a set of barriers determines certain splitting of the leaves, we
can express the above quantity in terms of the resulting
groups. Suppose $\widehat{\mathcal{C}}=\{\widehat{C}_1, ...,
\widehat{C}_M\}$ is a splitting of the leaves $[p]$, and $\mathcal{C}^*=\{C^*_1, ..., C^*_K\}$ is the true splitting. For each true group $C^*_i,\  i\in \{1,..., K\}$, we count the number of splits of $C^*_i$ by members of $\widehat{\mathcal{C}}$, i.e., $\sum_{j=1}^M \mathbbm{1}\{C^*_i \cap \widehat{C}_j \neq \emptyset\} - 1$. 
Therefore, the total number of excessive (false) splits of $C^*_i$ is given by
\[\sum_{i=1}^K \left( \sum_{j =1}^M \mathbbm{1}\{C^*_i \cap \widehat{C}_j \neq \emptyset\} -1\right)
= \sum_{i=1}^K \left( \sum_{j =1}^M \mathbbm{1}\{C^*_i \cap \widehat{C}_j \neq \emptyset\} \right)- K\,,\]
while the total number of splits is $(M-1)\vee 1$.
We define the {\em false split proportion} (FSP) and  true positive
proportion (interchanging $\mathcal{C}^*$ and $\widehat{\mathcal{C}}$) as 
\begin{equation}\label{eq:FSP-TPP} 
\small\FSP \coloneqq \frac{\sum_{i=1}^K \left(\sum_{j=1}^M \mathbbm{1}\{C^*_i \cap \widehat{C}_j \neq \emptyset\}\right)-K}{(M-1)\vee1}\,,\quad \TPP:= 1-\frac{\sum_{i=1}^M \left(\sum_{j=1}^K \mathbbm{1}\{C^*_i \cap \widehat{C}_j \neq \emptyset\}\right)-M}{K-1}\,.
\end{equation}
In the next lemma, we prove that the quantities $\FSP$ and $\TPP$ in terms of groups are equivalent to quantities $\FDP^{b}$ and $\TPP^{b}$ for the barrier discovery problem.
\begin{lemma}\label{lemma:FSRpowervsFDRpower}
For the quantities ${\rm FSP}$ and ${\rm TPP},$ given by \eqref{eq:FSP-TPP},  and the quantities ${\rm FDP}^{\rm b}$
and ${\rm TPP}^{\rm b}$,  given by \eqref{eq:FDPb-TPPb}, the following holds true:
${\rm FSP} = {\rm FDP}^{\rm b}\,, \quad {\rm TPP} = {\rm TPP}^{\rm b}$. 
\end{lemma}
We refer to Appendix~\ref{proof:lemma:FSRpowervsFDRpower} for the proof of Lemma~\ref{lemma:FSRpowervsFDRpower}.
The {\em false split rate} ($\FSR$) and the expected power are defined as
\begin{equation}\label{FSR} 
\FSR := \mathbb{E}(\FSP),\quad \text{Power} := \E(\TPP) \,,
\end{equation}
where the expectation is with respect to the randomness in $\widehat{\mathcal C}$, which in our context will depend on the $p$-values for the
hypotheses of the form \eqref{eq:HT-tree}.
In the next
section we provide another characterization for $\FSR$ in the tree-aggregation context, and in Section \ref{section:Testing} we develop a testing procedure that controls $\FSR$ at a pre-specified level $\alpha<1$.

\subsection{FSR on a tree} \label{section:FSR=FDR}
While the FSR metric can be calculated for a general splitting of $p$ objects using definition~\eqref{eq:FSP-TPP},
in this section we focus on splittings that can be expressed as a combination of branches of $\mathcal{T}$ as explained in the previous section. 
We will provide an equivalent characterization of $\FSP$ in this context in terms of specific structural properties of $\mathcal{T}$.  

For a testing procedure satisfying Constraint~\ref{cons1}, the rejected nodes
on the tree still maintain the tree structure. We use
$\mathcal{T}_{\rej}$ to represent the subtree of rejected nodes on the
tree $\mathcal{T}$.  We also define $\deg_{\mathcal{T}}(u)$ as the
(out) degree of node $u$ on tree $\mathcal{T}$ (the number of children
of node $u$); similarly, $\deg_{\mathcal{T}_{\rej}}(u)$ is the degree
of node $u$ on the subtree $\mathcal{T}_{\rej}$.  We use $\mathcal{F}$
as the set of false rejections in $\mathcal{T}$. Lastly, we define
$\mathcal{B}^*$ as the set of nodes whose leaf sets correspond to the
true aggregation, i.e., $\mathcal{B}^*$ is such that
$
\mathcal{C}^* = \{\mathcal{L}_u\  |\  u \in \mathcal{B}^*\}.
$
This characterization of $\mathcal{C}^*$ stems from the assumption that the true
aggregation is among the partitions allowed by the tree.

Our next lemma characterizes the number of false splits and the total number of splits in terms of the tree $\mathcal{T}$ and its subtree $\mathcal{T}_{\rej}$. By virtue of this lemma we have an alternative characterization of $\FSP$ (and $\FSR$), which is more amenable to analysis.

\begin{lemma}\label{lemma:FSP=V/R}
Define $V$ and $R$ as follows:
\begin{equation} \label{V-R} 
\small V \coloneqq \sum_{u \in \mathcal{F}} \left(\deg_{\mathcal{T}}(u) -\deg_{\mathcal{T}_{\rej}}(u)\right) - \left| \mathcal{B}^* \cap \mathcal{F}\right|,~~R \coloneqq \max \left\{\sum_{u \in \mathcal{T}_{\rej}} \left({\deg}_{\mathcal{T}}(u) - {\deg}_{\mathcal{T}_{\rej}}(u) \right)  -1, 0\right\}.
\end{equation}
Then $V$ and $R$ quantify the number of false splits and the total number of splits,  respectively. 
Consequently, we have
$\text{\rm FSP} = V/R$ and $\text{\rm FSR} = \mathbb{E}\left( V/R\right),$
where $\text{\rm FSP}$ and $\text{\rm FSR}$ are defined as in \eqref{eq:FSP-TPP} and \eqref{FSR}.
\end{lemma}

\begin{figure}
	\centering
	\includegraphics[width=0.6\textwidth]{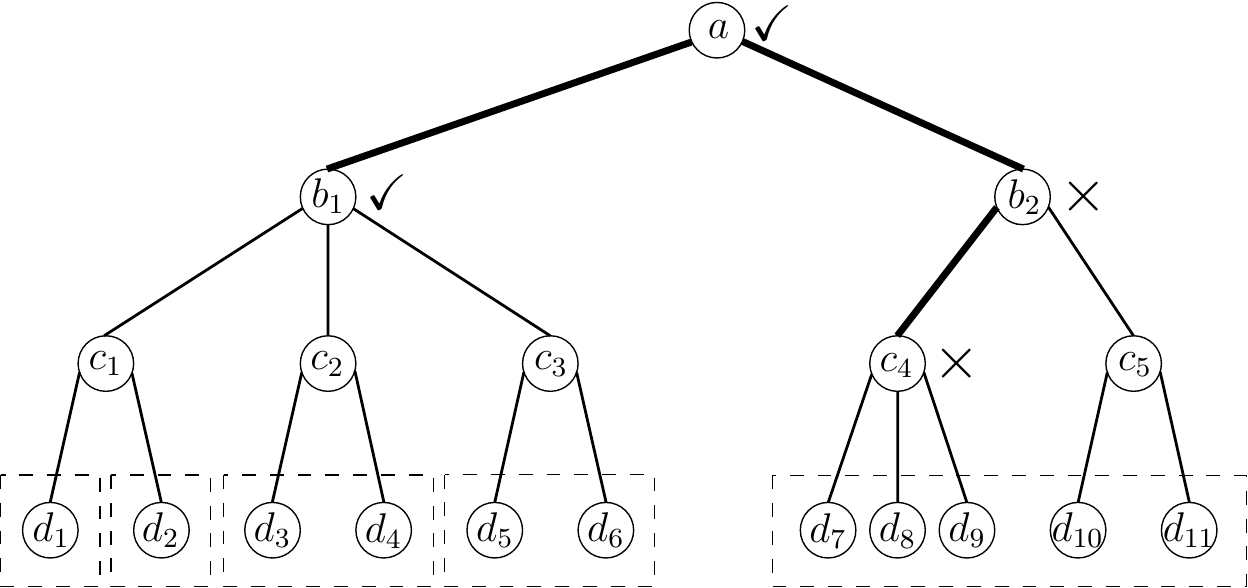}
	\caption{\em\small An example of $\mathcal{T}$, in which dashed boxes show the true aggregation
          of the leaves, $\mathcal C^*$, into $K=5$ groups, with
          $\mathcal B^*=\{d_1,d_2,c_2,c_3,b_2\}$.  The thicker edges
          and the nodes they connect form
          $\mathcal{T}_{\textrm{rej}}$, with $\checkmark$'s
          marking true rejections and $\times$'s marking false rejections
          $\mathcal{F}$.  The rejections correspond to an
          achieved aggregation with $M=7$ groups: $\{d_1, d_2\}, \{d_3,
          d_4\}, \{d_5, d_6\}, \{d_7\}, \{d_8\}, \{d_9\},  \{d_{10},
          d_{11}\}$. On the right branch of the tree, two false
          rejections lead to a nonzero false split rate, $\FSP =
          \frac{8 - 5}{7-1} = \frac{3}{6}$. We have $V = (3-0)+(2-1)-1
          = 3$, and $R =(2-2)+ (3-0)+ (2-1) + (3-0) -1 = 6$. Hence
          $\frac{V}{R} = \FSP$.  On the left branch of the tree, there
        is one missing rejection ($c_1$) that leads to a true
        positive proportion of $\TPP=1-\frac{8-7}{5-1}=\frac{3}{4}$.}
	\label{fig:example_FSR}
\end{figure}

A key quantity in the above characterization is $\deg_{\mathcal{T}}(u)
-\deg_{\mathcal{T}_{\rej}}(u)$, which counts the number of additional
splits due to rejecting $\hnull{u}$.
Figure~\ref{fig:example_FSR} represents a concrete example to illustrate the quantities and the equivalence stated in the lemma. 

\begin{remark}
Let us stress that the {\FSP} metric in general can be very different
from the standard~\FDR~metric for multiple hypothesis testing.
\FDR~measures the overall performance of the testing rule, including the hypotheses at the inner nodes, while \FSR~concerns the quality of the splitting of the leaves.
Therefore, methods for controlling \FDR~on trees cannot be applied to
control \FSR (as shown numerically in Section~\ref{sec:simulation_idealized_nonbinary_tree}). That said, we show in the next lemma that \FSP~and \FDP~become equivalent for the special case of a binary tree.   
\end{remark}
The following corollary states the equivalence for the special case in
which $\mathcal{T}$ is a binary tree. In this case, \FSP~corresponds
exactly to the commonly used \FDP, which is the ratio between the
number of false rejections and the total number of rejections.

\begin{lemma}\label{coro:binaryFSP=V/R}
	
For a binary tree,  the quantities $V$ and $R$ given by \eqref{V-R} can be simplified as
	$V = \left|\mathcal{F}\right|$ and
	$R = \left|\mathcal{T}_{\rm rej}\right|$.
	Therefore, 
	$\text{\rm FSP} =
        \left|\mathcal{F}\right|/\left|\mathcal{T}_{\rm rej}\right|$
        and $\FSR = \FDR:= \E\left(|\mathcal{F}|/|\cT_{\rm rej}|\right)$.
\end{lemma}
We defer the proofs for Lemma \ref{lemma:FSP=V/R} and Lemma~\ref{coro:binaryFSP=V/R} to Appendix \ref{proof:lemmas}.

\section{Hierarchical aggregation testing with FSR control}\label{section:Testing}

So far we have defined the metric $\FSR$ to measure the quality of a
splitting of leaves and proposed an alternate characterization of it
in terms of the structure of the rejected (and false rejected) nodes
as in Lemma~\ref{lemma:FSP=V/R}.  In this section, we introduce a new multiple testing
procedure to test the null hypotheses $\hnull{u}$, starting from the
root and proceeding down the tree.
The procedure assumes that each non-leaf node $u$ has a $p$-value that
is super-uniform under $\hnull{u}$, i.e.
\begin{align}\label{eq:supUnif}
\prob(p_u\le t) \le t \quad \text{ for all } t\in[0,1]\,.
\end{align}
Later, in Section \ref{section:two_applications}, we discuss how to construct such $p$-values for two statistical applications.

We call our multiple testing procedure $\HAT$,
shorthand for \emph{hierarchical aggregation testing}, as the 
parameters in the returned splits can be aggregated together to
improve model interpretability and in some cases improve the
predictive power of the model. The $\HAT$ procedure
controls the $\FSR$ both for independent $p$-values (Section
\ref{sec:hat-indep}) and under arbitrary dependence of the $p$-values
(Section \ref{sec:hat-arbitrary}).

The hypotheses defined in \eqref{eq:HT-tree} are indeed intersection
hypotheses, i.e.,
\begin{align}\label{eq:nested}
\hnull{u}\ \text{holds}\ \Rightarrow \hnull{v}\ \text{holds for}\  \forall v\in \mathcal{T}_u,
\end{align}
where $\mathcal{T}_u$ is the subtree rooted at node $u$. In other
words, the parent of a non-null node must be non-null, and if a node
is null then every child of it is null as well. This property
motivates us to use a top-down sequential testing algorithm on the
tree that honors Constraint \ref{cons1}.

Before describing the $\HAT$ algorithm, we establish some notation. 
We sometimes write $\hnull{d, u}$ to make it
explicit that node $u$ is at depth $d$ of the tree,
where the depth of a node is one plus the length of the unique path
that connects the root to that node (the root is at depth 1). We also use
$\mathcal{T}^d$ for the set of non-leaf nodes at depth $d$ of $\mathcal{T}$. 

The testing procedure runs as follows. Let $\alpha$ be our target FSR level. Starting from the root node, at each level $d$ we only test hypotheses at the nodes whose parents are rejected. The test levels for hypotheses are determined by a  step-up threshold function so that the test level at each hypothesis $\hnull{d, u}$ depends on the number of leaves under this node $|\mathcal{L}_u|$, the target level $\alpha$, the maximum node degree denoted by $\Delta$, and the number of splits made in previous levels, denoted by $R^{1:(d-1)}$.
The details of our $\HAT$ procedure are given in
Algorithm~\ref{tree algo}, and depend on node-specific thresholds
$\alpha_u(r)$, both explicitly and through the function
\begin{align}
R^d(r) : = \sum_{u\in\mathcal{T}^d} \ind\{p_u\le \alpha_u(r)\}
  (\deg_{\mathcal{T}}(u)-1).\label{eq:Rd-def}
\end{align}

\begin{algorithm}
	
	\begin{algorithmic}[1]
		\caption{\em\small Hierarchical Aggregation Testing $(\HAT)$ Algorithm}\label{tree algo}
		
		\REQUIRE: FSR level $\alpha$, Tree $\mathcal{T}$,
                $p$-values $p_u$ for $u\in\mathcal T\setminus \mathcal L$.
		\ENSURE: Aggregation of leaves such that the procedure controls FSR.
		
		\textbf{initialize} $\mathcal{T}_{\rm{rej}}^1 = \left\{\text{root}\right\}$, $R^{1:1} = \deg_\cT(\text{root})-1$.
		\REPEAT
		\STATE From depth $d=2$ to maximum depth $D$ of the tree $\mathcal{T}$, perform hypothesis testing on each node in $\mathcal{T}^d$. 
		Compute $r^*_d$ as 
		\[
		r^*_d = \max \left\{r\ge 0:\;\;  r\le R^d(r)\right\}\,,
		\]
        where $R^d(r)$ is defined in \eqref{eq:Rd-def}, with threshold
        function $\alpha_u(r)$ given by \eqref{eq:generalthreshold}
        (for case of independent $p$-values) or
        \eqref{eq:threshold_reshaping} (under general dependence among $p$-values).
        Reject the nodes in the set $\mathcal{T}^{d}_{\rm rej}=\left\{u \in \mathcal{T}^d: p_u \leq \alpha_u(r^*_d) \right\}$. 
		\\
		
		\STATE Update $\mathcal{T}^{1:d}_{\rm rej} = \mathcal{T}^{1:(d-1)}_{\rm rej} \cup  \mathcal{T}^{d}_{\rm rej}$,
		and $R^{1:d} = R^{1:(d-1)} + r_d^*$.
		\UNTIL{No node in the current depth has a rejected parent or $d = D$.}
	\end{algorithmic}
	
\end{algorithm}

\subsection{Testing with independent $p$-values}\label{sec:hat-indep}

Assuming that the node $p$-values $p_u$ are independent, the threshold function $\alpha_u(r)$ used for testing $\hnull{d, u}$ is defined as:
\begin{align}\label{eq:generalthreshold}
\alpha_{u}(r)=\mathbbm{1}\{\parent(u)\in\mathcal{T}^{d-1}_{\rm rej} \} \;
   \frac{1}{\Delta}\; \frac{\alpha |\mathcal{L}_u|(R^{1:(d-1)}+r)}{p(1-\frac{1}{\Delta^2})\har_{d,r}+\alpha |\mathcal{L}_u|(R^{1:(d-1)}+r)}\,,
\end{align}
where $\har_{d,r}$ is the partial harmonic sum given by
\begin{align}
\har_{d,r} = 1+  \sum_{m= R^{1:(d-1)}+r+1}^{p-1-\left(\sum_{u\in\cT^d} \deg_\cT(u) - |\cT^d| - r\right)} \frac{1}{m}\,.
\end{align}

\begin{thm}\label{thm:generalFSRcontrol}
	Consider a tree with maximum node degree $\Delta$ and suppose that for each node $u$ in the tree, under the null hypothesis $\hnull{u}$, the $p$-value $p_u$ is super-uniform (see~\eqref{eq:supUnif}).
	Further, assume that the $p$-values for the null nodes are independent from each other and from the non-null $p$-values. Then using Algorithm~\ref{tree algo} with threshold function \eqref{eq:generalthreshold} to test intersection hypotheses $\hnull{u}$ controls $\FSR$ under the target level $ \alpha$. 
\end{thm}

The proof of Theorem~\ref{thm:generalFSRcontrol} is given in
Section~\ref{sec:proof:thm:generalFSRcontrol} of the appendix and uses a combination of different ideas. At the core of the proof is a `leave-one-out' technique to decouple the quantities $V$ and $R$. We also use the following \emph{self-consistency} property of the testing rule. 
Observe that $R^d(r)$ counts the additional splits of the leaves that result due to the rejected nodes in depth $d$, assuming that the threshold level $\alpha_u(r)$ is used.
We prove that the following self-consistency property holds: $R^d(r_d^*) = r^*_d$ where $r^*_d$ is defined in Step 2 of Algorithm \ref{tree algo}. In words, using $r^*_d$ to test the nodes in $\mathcal{T}^d$ (node $u$ to be tested at level $\alpha_u(r^*_d)$) gives us $r^*_d$ additional splits of the leaves, and therefore the update rule for $R^{1:d}$ in line 3 of the algorithm ensures that this quantity counts the number of splits formed from testing nodes in depth $1, \dotsc, d$. Using the self-consistency property and  the leave-one-out technique,  along with intricate probabilistic bounds in terms of structural properties of $\mathcal{T}$, we prove that $\FSR$ is controlled at the pre-assigned level $\alpha$.  

A few remarks are in order regarding the testing threshold $\alpha_u(r)$. From its definition, we have $\alpha_u(r)=0$ if the parent hypothesis of $u$ is not rejected. Also note that since the testing is done in a downward manner, the event $\{\parent(u)\in\mathcal{T}^{d-1}_{\rm rej}\}$ is observed by the time the node $u$ is tested. Also note that as we reject more hypotheses early on, the burden of proof reduces for the subsequent hypotheses, because $\alpha_u(r)$ is increasing in $R^{1:(d-1)}$. This trend is  similar to FDR control methods (e.g.,~\cite{BH1995,javanmard2018online}). We also observe that $\alpha_u(r)$ is increasing in $|\cL_u|$. For the nodes at upper levels of the tree, this is crucially useful as $R^{1:(d-1)}$ is small for these nodes, while $|\cL_u|$ is large and compensates for it in the threshold function.  

Our next theorem is a generalization of
Theorem~\ref{thm:generalFSRcontrol} to the case that the null
$p$-values distribution deviates from a super-uniform distribution. We
will use Theorem \ref{thm:generalFSRcontrol-ext} to control $\FSR$ in
Section~\ref{section:equality_regression_coeffs} where we aim to
aggregate the features in a linear regression setting. As we will discuss, for this application we suggest to construct the $p$-values using a debiasing approach, which results in $p$-values that are asymptotically super-uniform (as the sample size $n$ diverges).

\begin{thm}\label{thm:generalFSRcontrol-ext}
	Consider a tree with maximum node degree $\Delta$ and suppose
        that for each non-leaf node $u$ in the tree, under the null hypothesis $\hnull{u}$, the $p$-value $p_u$ satisfies
	$\prob(p_u\le t) \le t+\eps_0\, \text{ for all } t\in[0,1]$,
	for a constant $\eps_0>0$.
	Further, assume that the $p$-values for the null nodes are independent from each other and from the non-null $p$-values. 
	Consider running Algorithm~\ref{tree algo} to test intersection hypotheses $\hnull{u}$ with the threshold function given by
	\begin{align}
	\alpha_{u}(r)=\mathbbm{1}\{\text{\rm parent}(u)\in\mathcal{T}^{d-1}_{\rm rej} \} \;
   \left\{\frac{1}{\Delta}\; \frac{\alpha |\mathcal{L}_u|(R^{1:(d-1)}+r)}{p(1-\frac{1}{\Delta^2})\har_{d,r}+\alpha |\mathcal{L}_u|(R^{1:(d-1)}+r)} -\eps_0 \right\}\,.
\end{align}
	
	Then, $\FSR$ is controlled under the target level $\alpha$. 

\end{thm}

\subsection{Testing with arbitrarily dependent $p$-values}\label{sec:hat-arbitrary}

Theorems \ref{thm:generalFSRcontrol} and \ref{thm:generalFSRcontrol-ext} assume that the null $p$-values are independent from each other and from the non-null $p$-values. To handle arbitrarily dependent $p$-values, we propose a modified threshold function:
\begin{align}\label{eq:threshold_reshaping}
\alpha_{u}(r)=\mathbbm{1}\{\parent(u)\in\mathcal{T}^{d-1}_{\rm rej} \} \;
\frac{\alpha |\mathcal{L}_u|\cdot \beta_d(R^{1:(d-1)}+r)}{p(\Delta-\frac{1}{\Delta})(D-1)}\,,
\end{align}
where $\beta_d(\cdot)$ is a reshaping function of the form
\begin{align}\label{eq:beta_d}
\beta_d(R^{1:(d-1)} + r) = \frac{R^{1:(d-1)} + r}{\sum_{k=d(\delta-1)}^{\sum_{u \in \mathcal{T}^d} \deg_{\mathcal{T}} (u)} \frac{1}{k}},
\end{align}
and $\delta$ is the minimum node degree in $\mathcal{T}\setminus\L$. It is
straightforward to see that the reshaping function is lowering the
test thresholds compared to the independent $p$-values case, making
the testing procedure more conservative to handle general dependence
among $p$-values. In the next theorem, we show that with the reshaped
testing threshold $\FSR$ is controlled for arbitrarily dependent $p$-values.
\begin{thm}\label{thm:reshapeFSRcontrol}
	Consider a tree with maximum node degree $\Delta$ and minimum
        node degree $\delta$, and suppose that for each node $u$ in
        the tree, under the null hypothesis $\hnull{u}$, the $p$-value
        is super-uniform, i.e., \eqref{eq:supUnif} holds.
	The $p$-values for the nodes can be arbitrarily dependent. Then, $\HAT$ (Algorithm~\ref{tree algo}) with the reshaped threshold~\eqref{eq:threshold_reshaping} controls $\FSR$ under the target level $ \alpha$. 
\end{thm}

The proof of Theorem~\ref{thm:reshapeFSRcontrol} builds upon a lemma
from~\cite{Blanchard_2008}  on dependency control of a pair of
non-negative random variables. We refer to
Section~\ref{proof:thm:reshapeFSRcontrol} of the appendix for further details and the complete proof.

We conclude this section with an analogous result to Theorem~\ref{thm:reshapeFSRcontrol}, where the $p$-values are \emph{approximately} super-uniform. This can also be perceived as a generalization of Theorem~\ref{thm:generalFSRcontrol-ext} to the case of arbitrarily dependent $p$-values.    

\begin{thm}\label{thm:reshapeFSRcontrol-ext}
	Consider a tree with maximum node degree $\Delta$ and minimum
        node degree $\delta$, and suppose
        that for each non-leaf node $u$ in the tree, under the null hypothesis $\hnull{u}$, the $p$-value $p_u$ satisfies
	\[
	\prob(p_u\le t) \le t+\eps_0\,,\quad \text{ for all } t\in[0,1]\,,
	\]
	for a constant $\eps_0>0$.
	The $p$-values for the nodes can be arbitrarily
        dependent. Consider running Algorithm~\ref{tree algo} to test
        the hypotheses $\hnull{u}$ with threshold function given by
	\begin{align}\label{eq:threshold_reshaping_asym}
\alpha_{u}(r)=\mathbbm{1}\{{\rm parent}(u)\in\mathcal{T}^{d-1}_{\rm rej} \} \;
\left\{\frac{\alpha |\mathcal{L}_u|\cdot \beta_d(R^{1:(d-1)}+r)}{p(\Delta-\frac{1}{\Delta})(D-1)} -\eps_0\right\}\,,
\end{align}
	with the reshaping function $\beta_d(\cdot)$ defined by~\eqref{eq:threshold_reshaping}. Then, $\FSR$ is controlled under the target level $ \alpha$. 
\end{thm}

Proof of Theorem~\ref{thm:reshapeFSRcontrol-ext} is similar to the
proof of Theorem~\ref{thm:reshapeFSRcontrol}, and is deferred to
Section~\ref{proof:thm:reshapeFSRcontrol-ext} of the appendix.

\section{Two statistical applications}\label{section:two_applications}

Here we consider two statistical applications of tree-based
aggregation. In Section \ref{section:equality_means}, we study the
problem of testing equality of means, for which the nodewise
$p$-values are formed by one-way ANOVA tests. In Section
\ref{section:equality_regression_coeffs} we study the problem of
aggregating features with the same coefficients in a linear regression
setting.

\subsection{Testing equality of means}\label{section:equality_means}
In this application, we imagine that $\btheta^*$ is a vector of
unknown means and that at each leaf node $i$ of a tree $\T$ there is a noisy
observation of the corresponding mean:
$y_i = \theta^*_i + \varepsilon_i$, where the $\varepsilon_i\sim
\normal(0,\sigma^2)$ are independent. 
Given the $y_i$, we want to aggregate the leaves by testing the
equality of their means.
For each node $u\in\mathcal{T}$, we construct a $p$-value based on a
one-way ANOVA test with known $\sigma>0$,
\begin{equation}
  \label{eq:anova-pvalue}
  p_u=1-F_{\chi^2_{\Delta_u-1}}\left(\sigma^{-2}\sum_{v\in\child(u)}|\cL_v|(\bar
    y_v - \bar y_u)^2\right),
\end{equation}
where $\bar y_v=|\cL_v|^{-1}\sum_{i\in \L_v}y_i$, and $\child(u)$ is the set of children of $u$. Also $\Delta_u:= \deg_{\mathcal{T}}(u) = |\child(u)|$ and $F_{\chi^2_{\Delta_u-1}}$ is the cdf of a
$\chi^2_{\Delta_u-1}$ random variable. 
We show in the following lemma that the above construction gives bona fide $p$-values for our testing procedure.
\begin{lemma}\label{lemma:ANOVA}

The $p$-value defined in \eqref{eq:anova-pvalue} is uniform under
$\H^0_u$ in \eqref{eq:HT-tree}. Furthermore, for any two distinct nodes $a,b\in \T\setminus\L$, $p_a$ and $p_b$ are independent. 
\end{lemma}

Recall that the nodewise hypotheses $\{\hnull{u}\}_{u\in\T\setminus\L}$ are intersection
hypotheses as in \eqref{eq:nested}, and therefore one can apply Simes'
procedure to form bona fide intersection $p$-values.

The Simes' $p$-value at node $a$ is given by 
$p_{a, \textrm{Simes}}:= \min_{1 \leq k \leq |\T_a\setminus\L_a|} \left(p_{(k)} \cdot |\T_a\setminus\L_a|\right)/k,$
where $p_{(k)}$ is the $k$th smallest $p$-value in $\T_a\setminus\L_a$.
As shown by \cite{Simes1986}, as the original $p$-values are
independent (as per Lemma~\ref{lemma:ANOVA}), the Simes' $p$-values
constructed as above are super-uniform, and hence can be used to test
the nodewise hypotheses. However, note that the Simes' $p$-values are
not independent anymore, so when applying the $\HAT$ procedure, we need to use the reshaped threshold function \eqref{eq:threshold_reshaping}.
 
\subsection{Testing equality of regression coefficients}\label{section:equality_regression_coeffs}

Consider a linear model where the response variables are generated as 
$\by\sim\normal(\bX \btheta^*,\sigma^2\boldsymbol{I}_n).$

In many applications the features are counts data, i.e., $X_{ij}$
records the frequency of an event $j$ occurring in observation
$i$. \citet{Yan2018RareFS} note that when events rarely occur, a common practice is to remove the rare
features in a pre-processing step; however, they
show that when a tree is available, rare features can instead be aggregated to create informative predictors that count the
frequency of tree-based unions of events.
While \citet{Yan2018RareFS} focused on predictive performance, here we focus on
aggregation recovery itself by controlling \FSR. To do so, we use the point estimator of~\citet{Yan2018RareFS}, along with a debiasing approach to construct the nodewise $p$-values for our proposed testing procedure. 

The \cite{Yan2018RareFS}  point estimator is the solution to the optimization problem,
\begin{equation}\label{eq:RareObjective}
  \widehat{\btheta} \in\arg
  \min_{\btheta\in \mathbb{R}^p} \frac{1}{2n} \left\Vert\boldsymbol{y} - \boldsymbol{X}\btheta \right\Vert_2^2 + \min_{ \boldsymbol{\gamma} \in \mathbb{R}^{|\mathcal{T}|}} \;\; \lambda \left(\nu \sum_{u\in\cT\backslash{\rm root}} |\gamma_{u}|+ (1-\nu) \sum_{j=1}^{p}  |\theta_j|\right) \ \ \  \textrm{s.t.} \ \ \btheta = \boldsymbol{A\gamma}\,,
 \end{equation}
where $\boldsymbol{A} \in \mathbb{R}^{p \times |\mathcal{T}|}$ encodes
the tree structure with $A_{ij}$ indicating whether leaf $i$ is a descendant of node $j$.
The resulting $\widehat{\btheta}$ tends to be constant on
branches of the tree, leading to aggregated features.

\subsubsection{Constructing $p$-values for the null hypotheses}\label{section:$p$-value_section}
A challenge in constructing $p$-values for the null hypotheses $\hnull{u}$ given in
\eqref{eq:HT-tree} is that the  
distribution of the estimator $\widehat{\btheta}$ is not tractable. Moreover, due to the regularization
term, this estimator is biased. We therefore use a debiasing approach.

The debiasing approach was pioneered in
\citet{CI_Adel2013,Zhang2014,vandergeer,javanmard2018debiasing} for
statistical inference in high-dimensions where the sample size is much
smaller than the dimension of the features (i.e., $n\ll
p$). Regularized estimators such as the lasso
\citep{tibshirani1996regression} are popular point estimators in these
regimes however they are biased. 
The focus of the debiasing work has been on
statistical inference on individual model parameters, namely
constructing $p$-values for null hypotheses of the form
$\mathcal{H}_{0,i}: \theta^*_{i} = 0$. The debiasing approach has been
extended for inference on linear functions of model
parameters~\citep{cai2017confidence,cai2019optimal} and also general
functionals of them~\citep{javanmard2020flexible}. The original
debiasing method can also be used to perform inference on a group of
model parameters, e.g. constructing valid $p$-values for null
hypothesis $\mathcal{H}_{0}: \btheta_A = 0$  where the group size
$|A|$ is fixed as $n,p\to \infty$ (see e.g,~\citet[Section
3.4]{CI_Adel2013}). More recently, \citet{guo2019group} have studied
the group inference problem for linear regression model by considering
sum-type statistics. Namely, by considering quadratic form hypotheses,
$\mathcal{H}_0: \btheta_A^\top \bG \btheta_A=0$, for a positive definite
matrix $\bG$. They propose a debiasing approach to directly estimate
the quadratic form $\btheta_A^\top \bG \btheta_A$ and to provide asymptotically valid $p$-values for the corresponding hypotheses. 
The constructed $p$-values
are valid for any group size in terms of type-I error control.
This work also discusses how by a direct application of the methodology developed in~\citet{Meinshausen},
one can test significance of multiple groups, where the groups are defined
by a tree structure.   The method of~\citet{Meinshausen} is based on a hierarchical approach to test variables' importance. At the core, it constructs hierarchical adjusted $p$-values to account for the multiplicity of testing problems and controls the family wise error rate at the prespecified level. 
 At every level of the tree, the $p$-value adjustment is a weighted Bonferroni correction and across different levels it is a sequential procedure with no correction but with the constraint that if a parent hypothesis is not rejected then the procedure does not go further down the tree. By comparison, our $\HAT$ algorithm controls the FSR, a very different criterion than the family wise error rate. Also $\HAT$ does not do any adjustment to $p$-values, and instead chooses the threshold levels in a sequential manner depending on the previous rejections and the structural properties of the tree.

Here we follow the methodology of~\citet{guo2019group} to
construct valid $p$-values for the $\HAT$ procedure, using the point
estimator~\eqref{eq:RareObjective}. We write $\hnull{u}$ equivalently as
$\widetilde{\mathcal{H}}^0_u:  Q_u \coloneqq
\btheta_{\mathcal{L}_u}^{*\top} \bG_{u} \btheta^*_{\mathcal{L}_u} =0$,
where $\bG_u$ is the centering matrix and we use the shorthand
$\btheta_u:= \btheta_{\cL_u}$. 
To make inference on the quadratic form $Q_u$, we first consider the point estimator estimator $\widehat{Q}_u := \widehat{\btheta}_{u}^{\top} \boldsymbol{G}_{u} \widehat{\btheta}_{u}$, where $\widehat{\boldsymbol{\theta}}$ is the estimator given by \eqref{eq:RareObjective}. 
To debias $\widehat{Q}_u$ we first decompose the error term into
\[
\widehat{Q}_u - Q_u = 
\widehat{\btheta}_{u}^{\top} \boldsymbol{G}_{u} \widehat{\btheta}_{u} - {\btheta_{u}^*}^{\top} \boldsymbol{G}_{u} \btheta_{u}^* = 2\widehat{\btheta}_{u}^{\top}\boldsymbol{G}_{u}\left(\widehat{\btheta}_{u}  - \btheta_{u}^*\right) - \left( \widehat{\btheta}_{u} - \btheta_{u}^* \right)^{\top}\boldsymbol{G}_{u} \left( \widehat{\btheta}_{u} - \btheta_{u}^* \right).
\] 
The dominating term in this decomposition is
$2\widehat{\btheta}_{u}^{\top} \boldsymbol{G}_{u}
(\widehat{\btheta}_{u} - \btheta_{u}^*)$. The approach in \citet{guo2019group} is to develop an \emph{unbiased} estimate of this term and then subtract this estimate from $\widehat{Q}_u$. 
Given a projection direction $\widehat{\boldsymbol{b}}$, the unbiased
estimate is of the form
\[
\frac{1}{n}\widehat{\boldsymbol{b}}^{\top} \boldsymbol{X}^{\top} (\boldsymbol{y} - \boldsymbol{X}\widehat{\btheta}) = \widehat{\boldsymbol{b}}^{\top}\widehat{\boldsymbol{\Sigma}}  ( \btheta^* - \widehat{\btheta}) + \frac{1}{n} \widehat{\boldsymbol{b}}^{\top} \boldsymbol{X}^\top \beps,
\]
where $\widehat{\boldsymbol{\Sigma}} := \frac{1}{n}
\boldsymbol{X}^{\top} \boldsymbol{X}$. 
 The idea is to find a projection direction $\widehat{\boldsymbol{b}}$ such that $\widehat{\boldsymbol{b}}^{\top} \widehat{\boldsymbol{\Sigma}}  (\widehat{\btheta}- \btheta^*)$ is a good estimate for $\widehat{\btheta}_{u}^{\top} \boldsymbol{G}_{u} (\widehat{\btheta}_{u} - \btheta_{u}^*)$. 
The projection direction $\widehat{\boldsymbol{b}}$ is constructed by solving the following optimization problem:
\begin{gather}
\begin{aligned}
\widehat{\boldsymbol{b}} = & \arg \min_{\boldsymbol{b}} \ \boldsymbol{b}^{\top} \widehat{\boldsymbol{\Sigma}} \boldsymbol{b} \quad\text{s.t.} \  \  \max_{\boldsymbol{\omega} \in \mathcal{C}_u} \Big|\langle \boldsymbol{\omega}, \widehat{\boldsymbol{\Sigma}} \boldsymbol{b} - [\widehat{\btheta}^{\top}_{u}\boldsymbol{G}_u\ \  \boldsymbol{0} ]^{\top}\rangle\Big| \leq \| \boldsymbol{G}_u \widehat{\btheta}_{u} \|_2 \lambda_n\,,
\end{aligned}\label{eq:ConvexOptimization}
\end{gather}
where  $$\mathcal{C}_u =\left\{ \boldsymbol{e}_1, ..., \boldsymbol{e}_p,  \frac{1}{\| \boldsymbol{G}_u \widehat{\btheta}_{u}\|_2} [\widehat{\btheta}^{\top}_{u}\boldsymbol{G}_u\ \  \boldsymbol{0} ]^{\top}  \right\}$$
and $\lambda_n$ is chosen to be of order $\sqrt{\log(p)/n}$. 
Finally the debiased estimator for $Q_u$ is constructed as
$\widehat{Q}^{\rm d}_u := \widehat{\btheta}_{u}^{\top} \boldsymbol{G}_{u} \widehat{\btheta}_{u} + \frac{2}{n} \widehat{\boldsymbol{b}}^{\top} \boldsymbol{X}^{\top} (\boldsymbol{y} - \boldsymbol{X}\widehat{\btheta}).$
Suppose that the true model $\btheta^*$ is $s_0$ sparse (i.e., it has
$s_0$ nonzero entries). As shown in~\cite[Theorem 2]{guo2019group},
under the condition $s_0 (\log p)/\sqrt{n} \to 0$, and assuming that the initial estimator $\widehat{\btheta}$ satisfies $\|\widehat{\btheta}-\btheta^*\|_2\le C\sqrt{s_0(\log p)/n}$ and $\|\widehat{\btheta}-\btheta^*\|_1\le Cs_0\sqrt{(\log p)/n}$ for some constant $C>0$, then the residual $\widehat{Q}^{\rm d}_u - Q_u$ asymptotically admits a Gaussian distribution. More specifically, $\widehat{Q}^{\rm d}_u - Q_u = Z_u +\Delta_u$ where
\begin{align}\label{eq:noise}
{Z_u} \sim \normal(0,{\var}(\widehat{Q}^{\rm d}_u) ), \quad {\var}(\widehat{Q}^{\rm d}_u) = \frac{4\sigma^2}{n} \widehat{\boldsymbol{b}}^{\top} \widehat{\boldsymbol{\Sigma}} \widehat{\boldsymbol{b}} \,.
\end{align}
In addition, for any constant $c_1>0$, there exists a constant $c_2>0$ depending on $c_1$ such that 
\begin{align}\label{eq:bias}
\prob\left(|\Delta_u| \ge c_1 (\|\bG_u \widehat{\btheta}_u\|_2 + \|\bG_u\|_2)\frac{s_0\log p}{n}\right) \le 2pe^{-c_2n}\,,
\end{align}
The above bound state that with high probability the bias term $\Delta_u$ is of order $s_0 (\log p)/n$, while ${\var}(\widehat{Q}^{\rm d}_u)$ is of order $1/n$. Therefore under the condition $s_0 (\log p)/\sqrt{n} \to 0$ the noise term $Z_u$ dominates the bias term $\Delta_u$.\footnote{In~\cite{guo2019group}, the probability bound $pe^{-c_2n}$ was further simplified to $p^{-c'}$ since $n\gtrsim \log p$ and assuming $n,p\to \infty$.}

Note that ${\var}(\widehat{Q}^{\rm d}_u)$ involves the noise variance $\sigma^2$ (which is the same for all nodes $u$). Let $\widehat{\sigma}$ be a consistent estimate of $\sigma$. Then the variance of the debiased estimator $\widehat{Q}^{\rm d}_u$ is estimated by
\begin{equation} \label{eq:Qformula}
 \widehat{\var}_{\tau}(\widehat{Q}^{\rm d}_u) = \frac{4\widehat{\sigma}^2}{n}  \widehat{\boldsymbol{b}}^{\top} \widehat{\boldsymbol{\Sigma}} \widehat{\boldsymbol{b}} + \frac{\tau}{n} ,
\end{equation} 
for some positive fixed constant $\tau$. The term $\tau/n$ is just to ensure that the estimated variance is at least of order $1/n$ (in the case of $\widehat{\boldsymbol{b}}^{\top} \widehat{\boldsymbol{\Sigma}} \widehat{\boldsymbol{b}} =0 $), and so it dominates the bias component of
$\widehat{Q}^{\rm d}_u$.  The exact choice of $\tau$ does not matter in the large sample limit ($n\to\infty$).

Using this result, we construct the two-sided $p$-value for the null hypothesis $\widetilde{\mathcal{H}}^0_{u}$ as follows:
\[
p_u = 2\left[1- \Phi\left(\frac{|\widehat{Q}^{\rm d}_u|}{\sqrt{\widehat{\var}_\tau(\widehat{Q}^{\rm d}_u) }}\right)\right],
\]
where $\Phi$ is the cdf of the standard normal distribution. 
\begin{propo}\label{propo:p-val}
Consider the asymptotic distributional characterization of $\widehat{Q}^{\rm d}_u$ given by~\eqref{eq:noise} and \eqref{eq:bias}. 
Let $\widehat{\sigma} = \widehat{\sigma}(\by,\bX)$ be an estimator of
$\sigma$ satisfying, for any fixed $\eps>0$,
\[
\lim_{n\to\infty} \prob\Big(\Big|\frac{\widehat{\sigma}}{\sigma} - 1\Big|\ge \eps\Big) = 0\,.
\] 
Under the condition $s_0(\log p)/\sqrt{n}\to 0$, for any fixed arbitrarily small constant $\eps_0$ (say $0.001$), there exists $n_0>0$ such that for all $n>n_0$, $\prob(p_u\le t) \le t+ \eps_0$, for all $t\in[0,1]$.
\end{propo}
We refer to Appendix~\ref{proof:propo:p-val} for the proof of Proposition~\ref{propo:p-val}.
By virtue of Proposition~\ref{propo:p-val}, the constructed $p$-values
satisfy the assumption of Theorem~\ref{thm:reshapeFSRcontrol-ext} and therefore by running the $\HAT$ procedure we are able to control $\FSR$ under the target level.

\section{Simulations}\label{section:simulation_result}

In this section, we conduct simulation studies (using the {\tt simulator} R package \citealt{bien2016simulator}) to understand the
performance of $\HAT$ in different settings.

\subsection{Testing on a binary tree with idealized $p$-values}\label{subsection:simlation_idealized_binary_tree}

Since $\FSR$ and $\FDR$ are equivalent in the special case of a binary
tree (by Lemma~\ref{coro:binaryFSP=V/R}), we begin by comparing HAT
with a testing procedure proposed by \citet{lynch2016procedures} to
control $\FDR$ in the hierarchical testing context (For non-binary
trees there is no such reference to compare with, since $\FSR$ is  a
criterion proposed by the present work, and there is no other
algorithm in the literature to control $\FSR$). 
Their method, which we refer to as LG, corresponds to Algorithm
\ref{tree algo} with several modifications.  First, their thresholds
are given by
\begin{equation} \label{eq:LynchandGuo}
\alpha_u(r) = \alpha\frac{|\mathcal{L}_u(\widetilde{\mathcal T})|}{|\mathcal{L}_{\mathrm{root}}(\widetilde{\mathcal T})|}\frac{m_u(\widetilde{\mathcal T}) + R^{1:(d-1)}+ r -1}{m_u(\widetilde{\mathcal T})},
\end{equation}
where $\widetilde{\mathcal T}$ is the tree in which we take
$\mathcal T$ and remove the leaves, $m_u(\widetilde{\mathcal T})$ is the number of descendants of
node $u$ in $\widetilde{\mathcal T}$, $|\mathcal{L}_u(\widetilde{\mathcal T})|$ is the number of
leaves in $\widetilde{\mathcal T}$ that descend from $u$.  Also, they
initialize $R^{1:1}=1$ and, instead
of \eqref{eq:Rd-def}, they take
$
R^d(r)=\sum_{u \in \widetilde{\mathcal T}^d} \mathbbm{1}\left\{p_u \leq \alpha_u(r)\right\}.
$

We randomly generate $p$ points from ${{\sf Unif}}[0,1]$ and form a binary tree
structure among them using hierarchical clustering.
We let $K = |\mathcal{B}^*|$ be the number of true groups by cutting
the tree into $K$ disjoint subtrees with the \proglang{R} function
\code{cutree}.  The nodes that are the roots of the subtrees form
$\mathcal{B}^*$. All non-leaf nodes in $\mathcal{B}^*$ and their
non-leaf descendants are null nodes, and we generate their $p$-values
independently from ${\sf Unif}([0,1])$. All ancestors of
$\mathcal{B}^*$ are non-null nodes, with $p$-values we generate independently from $\text{\rm Beta}(1, 60)$. 

For each pair of $p$ and $K$, the set of $p$-values are simulated
independently for 100 repetitions as described above.
We calculate $\FSP$ and $\TPP$ based on the aggregation of leaves that
results and average over the 100 values to estimate $\FSR$ and the mean power.

\begin{figure}
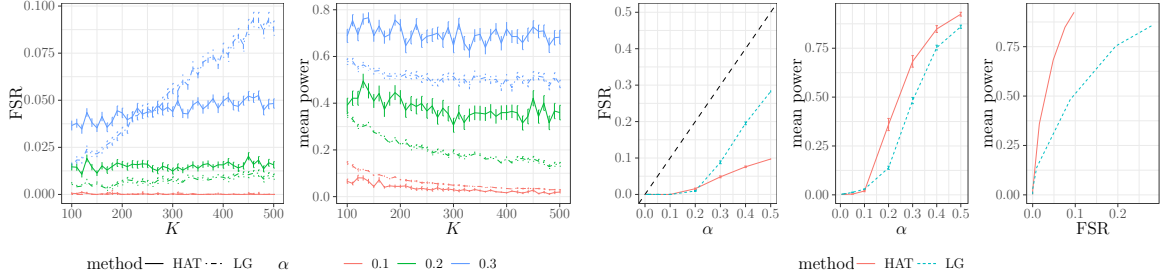

  \centering
  \scalebox{0.5}{\input{figures/binary_idealized_varying_k.tex}}
  \scalebox{0.5}{\input{figures/binary_idealized_varying_q.tex}}
	\caption{\em\small Plots of achieved FSR and average power by our
          algorithm (HAT) and Lynch and Guo's algorithm (LG), on a
          binary tree with $p=1000$ leaves and independent
          $p$-values. For the right three panels, $K=500$.}
	\label{fig:binary_idealized}
\end{figure}

The left two panels of Figure~\ref{fig:binary_idealized} show how $\FSR$ and
average power change with $K$ when $p$ is fixed at $1000$.  We can see
that both methods control $\FSR$ under the target $\alpha$'s. In terms
of power, when $\alpha=0.1$, the LG method enjoys slightly higher
power. For larger $\alpha$, however, the average power achieved by our
$\HAT$ method is higher; the gap in power enlarges as $K$
increases. When $K$ is large with the tree fixed, meaning that the
$\mathcal{B}^*$ nodes are at deeper levels, LG's power drops at a
faster rate than ours. Indeed, for these $\alpha$ values, our method
shows a substantial advantage when we have a deep tree and the non-null nodes appear at deeper levels of the tree.

The right three panels of Figure~\ref{fig:binary_idealized} show how achieved $\FSR$ and average power change with $\alpha$ in the setting where $p=1000, K=500$. We observe again that $\HAT$ achieves higher power than LG when $\alpha$ is above 0.1. From the left panel, we see that both methods are conservative in that the achieved $\FSR$ is lower than the target level $\alpha$, but as evident from the right-most panel, $\HAT$ showcases a better tradeoff between $\FSR$ and the mean power.

\subsection{Testing on a non-binary tree with idealized $p$-values}\label{sec:simulation_idealized_nonbinary_tree}
The LG algorithm is guaranteed to control $\FSR$ in the previous section due to the equivalence between $\FSR$ and $\FDR$ in the special case of a binary tree. However, for a non-binary tree, the LG algorithm does not have a theoretical guarantee on $\FSR$ control.

\begin{figure}
	\centering
	\scalebox{0.6}{\input{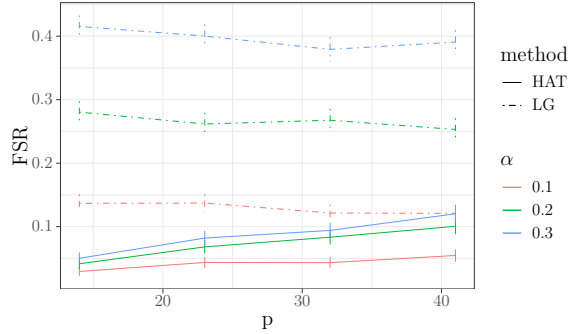}}
	\caption{\em\small Plot of achieved FSR by HAT and LG on a
          non-binary tree with $K=5$ and independent $p$-values. LG
          does not control $\FSR$ under the target levels.}
	\label{fig:nonbinary}
\end{figure}

We generate a tree where the root has degree $5$, and each
  child of the root is either a  non-leaf node with degree $10$ or is
  a leaf node; we vary the number of non-root non-leaf nodes from $1$
  to $4$, which results in $p$ ranging from $14$ to $41$. The number
  of true groups is fixed at $5$, therefore the root is the only
  non-null node. We simulate $p$-values for the interior nodes in the
  same fashion as in
  Section~\ref{subsection:simlation_idealized_binary_tree}: the
  $p$-values for null nodes are simulated independently from ${\sf
    Unif}([0,1])$ and the $p$-values for non-null nodes are simulated
  independently from $\text{\rm Beta}(1, 60)$.
  An estimate of $\FSR$ is obtained by averaging $\FSP$ over 100 runs. The achieved $\FSR$ is shown in Figure~\ref{fig:nonbinary}. As expected, we observe that the $\HAT$ procedure controls FSR under each target $\alpha$ for all values of $p$, whereas the LG algorithm does not.

Therefore, for aggregating leaves in general settings where the tree
can be beyond binary, only our algorithm provably controls $\FSR$
under the pre-specified level. This highlights the importance of using
our approach, which has guaranteed $\FSR$ control for tree-based
aggregation problems with non-binary trees.

\subsection{Two statistical applications}\label{sec:sim-applications}

\subsubsection{Testing equality of means}
In this section we apply the $\HAT$ procedure to the problem of testing equality of means. To simulate this setting, we form a balanced 3-regular  tree with $p=243$ leaves. For each $K$, we cut the tree into $K$ disjoint subtrees, which leads to $K$ non-overlapping subgroups of leaves. We assign a value to each leaf as
$y_i = \theta^*_{k(i)} + \eps_i, \;\; k(i) \in \{1, ..., K\}, i \in \{1, ..., p\},$
where $k(i)$ represents the group of leaf node $i$ and the elements of
$\btheta$ are independently generated from a ${\sf Unif}(1, 1.5)$
distribution multiplied by random signs, and $\eps_i$'s from a $\normal(0, \sigma^2)$ distribution.
We simulate $100$ runs by generating $100$ independent $\beps$'s with the noise level set to $\sigma=0.3$. The $p$-values are calculated as in \eqref{eq:anova-pvalue}.

By Lemma \ref{lemma:ANOVA}, the ANOVA $p$-values are
independent. Thus, by Theorem \ref{thm:generalFSRcontrol}, we can
perform $\HAT$ using the  using threshold function \eqref{eq:generalthreshold}. Alternatively, we can form the bona fide $p$-value using Simes' procedure, and test with the reshaped threshold function that is designed for arbitrarily dependent $p$-values.

\begin{figure}
	\centering
	\scalebox{0.7}{\input{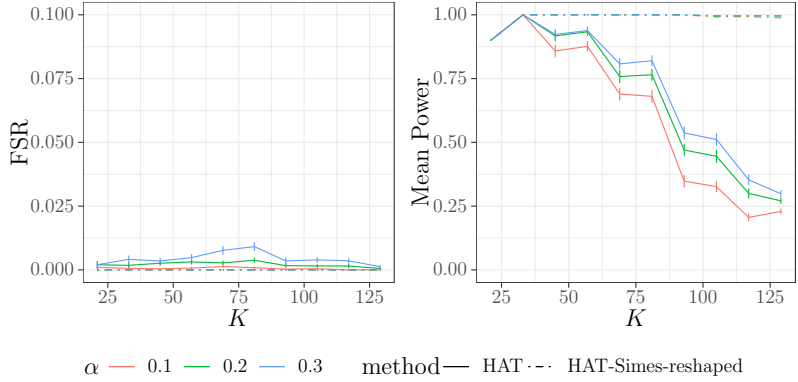}}
	\caption{\em\small Plots of achieved FSR and mean power with
          ANOVA $p$-values on a 3-regular tree ($p=243, \sigma =
          0.3$).}
	\label{fig:ANOVA_3_varying_k}
\end{figure}

We calculate $\FSR$ and average power by taking the average of the $\FSP$ and power over the $100$ runs. Figure~\ref{fig:ANOVA_3_varying_k} demonstrates how $\FSR$ and average power change with $K$. We observe that using Simes' $p$-values together with the reshaped thresholds achieves both lower FSR and higher power, which makes sense in this context because large effect sizes low in the tree may not translate to large effect sizes high in the tree.

\subsubsection{Testing equality of regression coefficients}
We apply $\HAT$ to the application of testing equality of regression
coefficients. We assume a high-dimensional linear model as described
in Section \ref{section:equality_regression_coeffs} and generate $p$
coefficients that take $K$ unique values. This partition comes from
leaves of disjoint subtrees of $\mathcal T$. We compute the $p$-values
using the debiased method on each node as in Section
\ref{section:$p$-value_section}. The details of the data generating
process are described in Section \ref{sec:simulation-regression} of
the appendix.

For each $K$, we simulate 100 independent $\beps$'s. The initial estimator $\widehat{\btheta}$ that solves the optimization problem~\eqref{eq:RareObjective} is achieved by using the \proglang{R} package \code{rare}~\cite{RarePackage}. The tuning parameters $\lambda$ and $\nu$ are chosen by cross-validation over a $2 \times 10$ grid.  We then follow the steps described in Section~\ref{section:$p$-value_section} to compute the $p$-values at each node. The positive constant $\tau$ in \eqref{eq:Qformula} is set to one and 
the noise level estimate $\widehat{\sigma}$ is obtained using the scaled lasso~\cite{sun2011scaled} (\proglang{R} package \code{scalreg}).

\begin{figure}
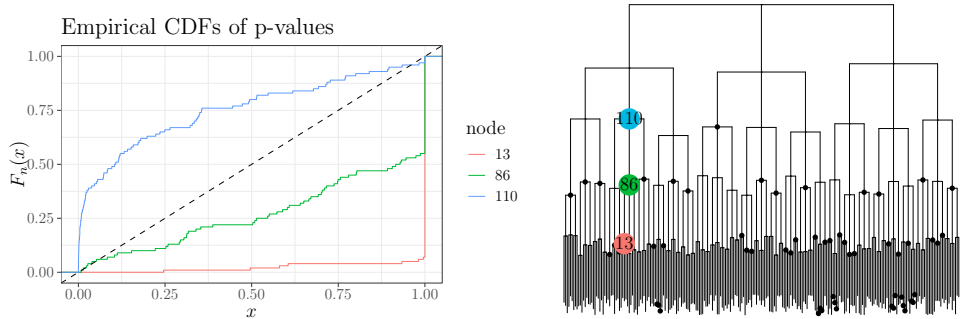

	\centering
	\scalebox{.55}{\input{figures/three_nodes_cdf.tex}}
        \scalebox{.45}{\input{figures/tree_with_three_nodes.tex}}
	\caption{\em\small Plots of empirical CDFs of three nodes under the
          setting $n=100$, $p = 243$, $\beta = 0.6$, $K = 30$, $\rho =
          0.2$, $\sigma = 0.6$. Node \#110 is a non-null node, node
          \#86 is a null node in $\mathcal B^*$, and node \#13 is a
          null node that is a child of \#86.}
	\label{fig:three_nodes_cdf}
\end{figure}

Figure~\ref{fig:three_nodes_cdf} shows the empirical cdf of the $p$-values, obtained from the 100 realizations of the noise, at three representative nodes when $K =57$. Among the three nodes, node \#110 is a non-null node, which means $\btheta^*_{\mathcal{L}_{110}}$  contains at least two distinct values.  Nodes \#13 and \#86 are both null nodes but at different depths on the tree. node \#86 is one of the $\mathcal{B}^*$ nodes and node \#13 is a descendant of node \#86. The curve of $p$-values at node \#110 is  above the diagonal line, which means the distribution has a higher density at small values than uniform distribution. On the contrary, the distribution of $p$-values at nodes \#13 and \#86 are super-uniform. The curve for a deeper level node seems to be further away from the diagonal line than its ancestor node.

\begin{figure}
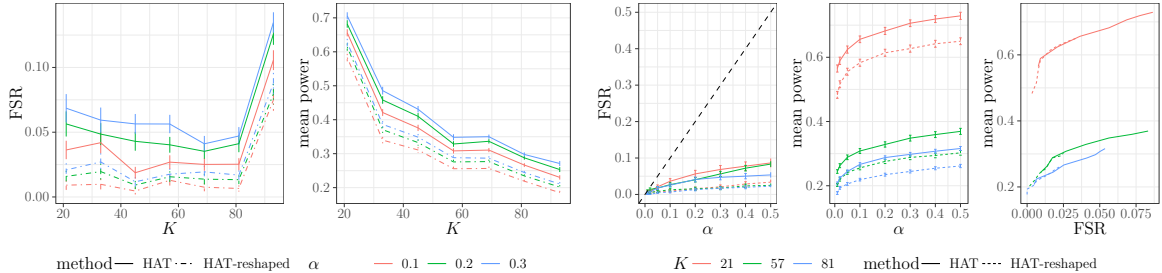

	\centering
	\scalebox{0.5}{\input{figures/debiased_regular_3varying_k.tex}}
	\scalebox{0.5}{\input{figures/debiased_regular_3varying_q.tex}}
	\caption{\em\small Plots of the achieved FSR and average power
          on a 3-regular tree ($n=100$, $p=243$, $\beta = 0.6$, $\rho
          = 0.2$, $\sigma = 0.6$) and $p$-values generated by the debiasing procedure.}
	\label{fig:debiased_regular_3}
\end{figure}

The $p$-values generated are not necessarily independent, so we use
the reshaped threshold function \eqref{eq:threshold_reshaping}, which
we have shown in theory controls $\FSR$ with arbitrarily dependent
$p$-values. We also test with the threshold
function~\eqref{eq:generalthreshold}, which we have not proven $\FSR$
control when the $p$-values are dependent. In Figure
\ref{fig:debiased_regular_3}, we demonstrate the result
for both threshold functions, varying $K$ and $\alpha$.
We observe from the plots that testing with both threshold functions control $\FSR$ below each target level $\alpha$. The reshaping function makes the threshold more conservative, hence the power of the $\HAT$ test with the reshaping function is generally lower. 

\section{Data examples} \label{section:real_data}

\subsection{Application to stocks data}

In this section, we analyze whether volatility of stocks is similar if
companies are in similar categories.
We use daily stock price data from January 1, 2015 to December 31, 2019,
derived from the US Stock Database \copyright 2021 Center for Research in
Security Prices (CRSP), The University of Chicago Booth School of
Business \citep{CRSP}. Specifically, we wish to aggregate stocks in a
similar sector unless their volatility levels are significantly different. We use several criteria for screening stocks of interest: We only keep common stocks that are publicly traded throughout this entire period; we also avoid penny stocks that have prices under \$0.01 per share. After pre-screening, we have $n=2538$ stocks in total. Following \citet{Parkinson1980} and \citet{MARTENS2007181}, we use the high-low range estimator for the daily variance 
$v_{t} = \frac{1}{4 \log(2)} (\log(H_{t} )- \log(L_{t}))^2,$
where $H_{t}$ and $L_{t}$ are day $t$'s highest and lowest prices,
respectively. We take the average of $v_{t}$ throughout the 5-year
period as our estimate for the  volatility of each stock and log-transform the volatility to reduce skewness.

We combine this stock log-volatility data with company industry classification information provided by the Compustat database \citep{Compustat}. The classification system we use is the North American Industry Classification System (NAICS), an industry classification system that employs a six digit code: the first two digits designate the largest sector; the third, fourth, fifth and sixth digits designate the subsector, industry group, industry, and national industry, respectively. We use this hierarchy to construct a tree with the first six layers representing the digits and the last layer, namely the leaves, corresponding to the individual companies.

At every node on the tree, we acquire a $p$-value by performing an
$F$-test (Equation 8.4, \citealt{seberlee}), for testing equality of
the log-volatilities of all stocks within the subtree defined by this
node. We further apply Simes' procedure to the $p$-values. We use
$\HAT$ with the reshaped thresholds and $\alpha = 0.4$. The achieved
aggregation result is summarized in Table
\ref{table:aggregation_companies} in Section
\ref{sec:companies} of the appendix.

\begin{figure}
	\centering
	\scalebox{0.6}{\input{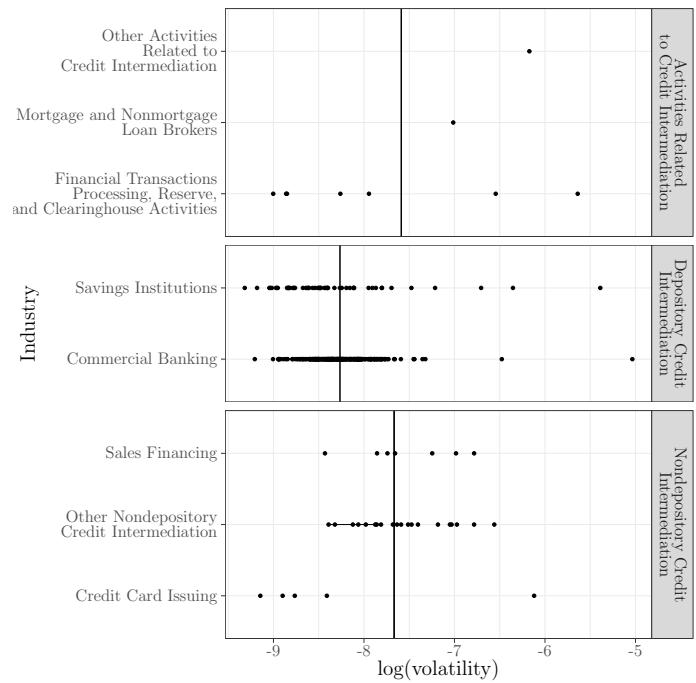}}
	\caption{\em\small The subsector ``Credit Intermediation and
          Related Activities" consists of $347$ companies, represented
          as points. These fall into 3 industry groups and 8 industries. Applying $\HAT$ rejects the null hypothesis that
          the 3 industry groups have the same mean $\log$-volatility,
          but does not reject this within each
          industry group.}
	\label{fig:subsector_credit}
\end{figure}

The final aggregation result consists of $40$ clusters at a variety of
levels: $21$ at sector level, $8$ at subsector level, $10$ at industry
group level, and one at company level. Two sectors ``Manufacturing II"
and ``Finance and Insurance" are split into further clusters while
other sectors remain undivided. Figure
  \ref{fig:subsector_credit} focuses on the $347$ companies in the
  subsector ``Credit Intermediation and Related Activities".  Each
  point represents the log-volatility of a company. The three facets
  correspond to three industry groups within the subsector and eight
  levels on the y-axis correspond to the eight industries nested in
  the industry groups. As can be observed in the plot, the industry
  group ``Depository Credit Intermediation" has significantly lower
  mean (around -8.27) compared to the other two industry groups in the
  subsector (around -7.67 and -7.59 respectively). Therefore, the null
  hypothesis that the three industry groups have similar mean
  volatility is rejected. On the contrary, within each industry group,
  there are no noticeable differences among different industries,
  leading none of the null hypotheses at the industry group level to be rejected.

\subsection{Application to New York City (NYC) taxi data}
\label{sec:taxi}

We apply our method of aggregating features to the NYC Yellow Taxi
Trip data (available at \url{data.cityofnewyork.us}),
restricting
attention to taxi trips made in December 2013. After cleaning the data, we
have 13.5 million trips made by $n=32704$ taxi drivers. We take
the total fare each taxi driver earned as the
response variable and take the number of rides starting from each
of $p=194$ neighborhood tabulation areas \citep{NTA} as the features.
We form a tree with NTAs as leaves, by connecting the root to
five nodes, representing the boroughs of NYC.  Within each
borough, we apply hierarchical clustering to the NTAs based on their
geographical coordinates. This results in a tree with depth 10.
The availability of taxis is not uniformly distributed across the city
(see Figure~\ref{fig:distribution_of_X} of Section
\ref{sec:taxis} of the appendix) and $\boldsymbol{X}$ is a highly
sparse matrix.

To aggregate neighborhood features, we perform the following
procedure: with data $\boldsymbol{X}$ and $\boldsymbol{y}$, as well as
the given tree structure, we first fit the penalized regression
\eqref{eq:RareObjective} to construct an initial estimate of the
coefficients $\widehat{\btheta}$. The estimation is achieved by using
the \code{rare} package with cross-validation across for choosing the
regularization parameters $\nu$ and $\lambda$ across a grid of $5
\times 50$ values. Next, we carry out the debiasing step by solving
the optimization problem~\eqref{eq:ConvexOptimization}, with the
\proglang{R} package \code{quadprog}. Note that the noise level
$\sigma$ is unknown, which we estimate by using the scaled
lasso (\citealt{sun2011scaled}; \proglang{R} package
\code{scalreg}). Moreover,  the positive constant $\tau$ in
\eqref{eq:Qformula} is set to one. After constructing the $p$-values
for each non-leaf node of the tree, we run $\HAT$ with $\alpha =
0.3$. 

\subsubsection{Aggregation results}
\label{sec:taxi-aggregation}

Our testing result 
leads to $44$
aggregated clusters, with the boroughs of Bronx and
Staten Island remaining undivided.
Brooklyn, Queens, and Manhattan are divided into $7$, $14$, and $21$ subgroups,
respectively. The left panel of Figure~\ref{fig:taxis}, we shows the
coefficients resulting from performing least squares on these $44$
aggregated features. Trips starting from Manhattan and parts of Queens, especially the
airports, have higher coefficient values. Within Manhattan,
areas like Hell's kitchen, Times Square, and Penn Station have some of
the higher coefficient values.

\begin{figure}
  \centering
  \begin{minipage}{0.5\linewidth}
\scalebox{.4}{\input{figures/mapwithcoeffs.tex}}
  \end{minipage}%
  \begin{minipage}{0.5\linewidth}
    \begin{tabular}{l|r }
Method & $\MSPE$\\
\hline
\hline
\code{L1} & 95.780\\
\code{L1-dense} & 95.864\\
\code{ls-boro} & 147.438\\
\code{L1-agg-h} & 102.593\\
\code{Rare} & 95.725\\
$\HAT$ (our method) & 95.459\\
\hline
\end{tabular}
  \end{minipage}
	\caption{\em\small  Left: Map colored with log-transformed least square
          coefficients from regressing fare on features from
          $\HAT$'s aggregation of neighborhoods of New York
          City. There are $44$ aggregated clusters out of the $194$
          neighborhoods. Darker colors correspond to higher fitted
          coefficients. Right: Prediction performance of the $6$ methods with the test data set.}
	\label{fig:taxis}
\end{figure}

In Section \ref{sec:changing-size} of the appendix we show, by taking
subsamples of different sizes, that reducing sample size leads to fewer rejections and therefore fewer aggregated groups.

\subsubsection{Comparing prediction performance}

In this section, we assess prediction performance achieved by our
aggregated features. We hold out a random sample of $20\%$ of the
drivers as the test set, and train with the remaining $80\%$. We
compare to the following models (each tuned via 10-fold cross validation):
\begin{itemize}
	\item Lasso with the original variables (\code{L1}).
	\item Lasso with only dense features (\code{L1-dense}): We
          drop features with $<0.5\%$ nonzeros then fit a lasso on the remaining $99$ features.
	\item Least squares with clusters aggregated to the five boroughs (\code{ls-boro}).
	\item Lasso with clusters aggregated at optimized height
          (\code{L1-agg-h}). We tune (over a grid of 5 values) an extra parameter $h$ that determines the
          aggregation height in the tree.
	\item Rare regression proposed by~\citet{Yan2018RareFS}
          (\code{Rare}). 
\end{itemize}

We compute the mean squared prediction error (MSPE) of each method on
the test set (see right panel of Figure~\ref{fig:taxis}). The \code{L1} and
\code{L1-dense} methods are not aggregation-related and achieve similar performance. Both \code{ls-boro} and \code{L1-agg-h} achieve some level of aggregation but the aggregations are determined at certain heights. \code{L1-agg-h} has an additional tuning parameter and is therefore advantageous. Lastly, both \code{Rare} and our method achieve aggregation in a flexible way, and the prediction results are comparable. \code{Rare} selects $43$ aggregation clusters while our method achieves $46$ groups in total.

In Section \ref{sec:taxis} of the appendix, we perform an additional
experiment with a synthetic response (but with $\boldsymbol{X}$ and
$\T$ from this data set) to measure the FSR and power.

\section{Conclusion} 
In many application domains, ranging from business and e-commerce, to computer vision and image processing, biology and ecology, the data measurements are naturally associated with the leaves of a tree which represents the data structure. Motivated by these applications, in this work we studied the problem of splitting the  measurements into non-overlapping subgroups which can be expressed as a combination of branches of the tree. The subgroups ideally express the  leaves that 
should be aggregated together, and perceived as single entities. 
We formulate the task of tree-based aggregation/splitting as a multiple testing problem and
introduced a novel metric called false split rate which corresponds to the fraction of splits made that were unnecessary. In addition, we proposed a so called $\HAT$ procedure (and a few variants of it) to return a splitting of leaves, which is guaranteed to control the false split rate under the target level. 
 
 It is worth noting some of the salient distinctions of the setup considered in this paper with the classical hierarchical clustering. Firstly, in hierarchical clustering the tree is cut at a fixed level, while our framework allows for more flexible summarization of the tree where different branches are cut at different depths. In other words, our framework yields multi-scale resolution of the data. Secondly, the clustering problem is often formulated as an unsupervised problem. In contrast, our framework can be perceived as supervised clustering problem where the labeled data are used to group the leaves by combining the branches of the tree.

\appendix
\section{Proof of main theorems} \label{section:proofs}
\subsection{Proof of Theorem \ref{thm:generalFSRcontrol}}\label{sec:proof:thm:generalFSRcontrol}

Recall the definition of the quantities $V$ and $R$:
\[
V := \sum_{u \in \mathcal{F}} \left(\deg_{\mathcal{T}}(u) -\deg_{\mathcal{T}_{\rej}}(u)\right) - \left| \mathcal{B}^*\cap \mathcal{F} \right|,
\]

\[
R := \max \left\{\sum_{u \in \mathcal{T}_{\rej}} \left({\deg}_{\mathcal{T}}(u) - {\deg}_{\mathcal{T}_{\rej}}(u) \right)  -1, 0\right\}.
\]

Note that $R>0$ because $\cT_{\rej} \subset \cT$ (recall that $\cT_{\rej}$ does not include any leaves of $\cT$ as there is no hypothesis associated to those nodes.) As we showed in Lemma~\ref{lemma:FSP=V/R}, the false split rate can be written in terms of $V$ and $R$:  
\[
\FSR = \mathbb{E}\left[ \frac{V}{R \vee 1} \right].
\]

For node $a\in \mathcal{B^*}$ let $\mathcal{F}_a = \mathcal{F} \cap \mathcal{T}_a$, and define the quantity $V_a$ as follows: 
\begin{align}\label{eq:Vaformula}
V_a=\begin{cases}
\sum_{u \in \mathcal{F}_{a}} \left(\deg_{\mathcal{T}}(u) - \deg_{\mathcal{T}_{a, \rm rej}}(u)\right) - 1\,,\quad &\text{ if } \mathcal{F}_a\neq \emptyset\\
0\, & \text{otherwise}
\end{cases}
\end{align}
By definition $V_a\ge0$. Indeed, from the proof of Lemma~\ref{lemma:FSP=V/R}, $V_a$ is the number of false splits in the set $\cL_a$. Also it is easy to verify that $V = \sum_{a\in \mathcal{B}^*} V_a$. 

We first show that
\begin{align}\label{eq:partialFSR}
\E\left(\frac{V_a}{R}\right)\le \frac{\alpha |\cL_a|}{p}\,, \quad \text{for } a\in \mathcal{B}^*.
\end{align}
Denote by $S(\mathcal{T}_a)$ the set of all nonempty subtrees of $\mathcal{T}_a$ rooted at node $a$. We also let $V_a(\Tp)$ be the number of false splits in $\cL_a$ when the rejection subtree is $\Tp$, i.e.,
	\[
	V_a(\Tp)=\sum_{u \in \Tp} \left(\deg_{\mathcal{T}}(u) - \deg_{\Tp}(u)\right) - 1\,.
	\]
	Here we used that $a\in \mathcal{B}^*$ and therefore any rejection in $\Tp$ is a false rejection and so $\mathcal{F}_a = \Tp$.
	Define $\tR_{\mathcal{T'}}$ to be the total number of splits when we set $p_u=0$ for $u\in \mathcal{T}'$ and $p_u = 1$ for $u \in\mathcal{T}_a\backslash \mathcal{T}'$.
	
	Note that $\tR_{\cT_{a, \rm rej}} = R$ since for $u\in \cT_{a, \rm rej}$ the $p$-value $p_u$ is already below the threshold at node $u$ and for $u\in \cT_a\backslash \cT_{a, \rm rej}$,  $p_u$ is already  above the threshold at that node $u$. 
	Therefore, writing $\mathcal{P}_{\mathcal{T}_a}^c = \{p_u: u\notin \cT_a\}$, we have 
	\begin{align}
	\E\left[\frac{V_a}{R}\ind(V_a>0)\bigg|\mathcal{P}_{\mathcal{T}_a}^c \right] &=\sum_{\mathcal{T'}\in S(\mathcal{T}_a)} \E\left[\frac{V_a(\cT')}{\tR_{\cT'}} \ind(\cT_{a, \rm rej} = \cT' )\bigg|\mathcal{P}_{\mathcal{T}_a}^c \right] \nonumber\\
	&= \sum_{\mathcal{T'} \in S(\mathcal{T}_a)} \frac{ V_a(\mathcal{T'}) }{\tR_{\cT'}} \cdot  \prob( \mathcal{T}_{a, \rm rej} = \mathcal{T'})\,, \label{eq:condFSR0}
	\end{align}
where  $S(\mathcal{T}_a)$ denotes the set of all nonempty subtrees of $\mathcal{T}_a$ rooted at node $a$, and we have used the fact that $V_a(\cT')$ is non-random and $\tR_{\cT'}$ is constant conditional on $\mathcal{P}_{\cT_a}^c$.
 
 Define $R^d(r) : = \sum_{u\in\mathcal{T}^d} \ind\{p_u\le \alpha_u(r)\} (\deg_{\mathcal{T}}(u)-1)$. Observe that $R^d(r^*_d)$  is the additional number of 
 splits made by the rejected nodes in depth $d$, going from depth $d-1$ to depth $d$, because the hypotheses $\hnull{u}$ in depth $d$ are tested at level $\alpha_u(r^*_d)$. Using our notation this can be written as the identity $R^{1:d} = R^{1:(d-1)} + R^d(r^*_d)$.
 
 We argue that $r^*_d = R^d(r^*_d)$.  To see why, note that by definition
 \[
 r^*_d = \max \left\{0\leq r \leq \sum_{u \in \mathcal{T}^d} \deg_{\mathcal{T}}(u)-|\mathcal{T}^d|:\;\;  r\le R^d(r)\right\}\,.
 \]
 Hence, $r^*_d\le R^d(r^*_d)$ and $r^*_d+1 > R^d(r^*_d+1)$. Since $R^d(r)$ is an integer valued function, the fact that $R^d(r^*_d+1)< r^*_d+1$ implies $R^d(r^*_d+1)\le r^*_d$. Thus, $r^*_d\le R^d(r^*_d)\le R^d(r^*_d+1) \le r^*_d$, which gives $r^*_d = R^d(r^*_d)$, and consequently
 \begin{align}\label{eq:R1d}
 R^{1:d} = R^{1:(d-1)} + r^*_d\,.
 \end{align}
 
 We next continue by upper bounding the right hand side of \eqref{eq:condFSR0}. Based on our testing methodology, described in Algorithm~\ref{tree algo}, a typical node $u$ at depth $d$ is tested at level $\alpha_u(r^*_d)$ given by~\eqref{eq:generalthreshold}. We have
 \begin{align}\label{eq:talpha}
 \alpha_u(r^*_d) &= \mathbbm{1}\{\parent(u)\in\mathcal{T}^{d-1}_{\rm rej} \} \;
   \frac{1}{\Delta} \; \frac{\alpha |\mathcal{L}_u|(R^{1:(d-1)}+r^*_d)}{p(1-\frac{1}{\Delta^2})\har_{d,r} +\alpha |\mathcal{L}_u|(R^{1:(d-1)}+r^*_d)}\nonumber\\
    &=  \mathbbm{1}\{\parent(u)\in\mathcal{T}^{d-1}_{\rm rej} \} \;
   \frac{1}{\Delta}\; \frac{\alpha |\mathcal{L}_u| R^{1:d}}{p(1-\frac{1}{\Delta^2})\har_{d,r} +\alpha |\mathcal{L}_u| R^{1:d}}\nonumber\\
     &= \mathbbm{1}\{\parent(u)\in\mathcal{T}^{d-1}_{\rm rej} \} \;
    \frac{1}{\Delta} \; \frac{\gamma_u}{p(1-\frac{1}{\Delta^2})+\gamma_u } \,,
 \end{align}
 with 
 $$\gamma_u:= \frac{\alpha}{\har_{d,r}} |\mathcal{L}_u| R^{1:d}\,.$$
 Note that $\alpha_u(r^*_d)$ is increasing in $\gamma_u$. 
 
 \begin{lemma}\label{lem:gamma-UB}
 Suppose that $u\in \cT_a$ and the node $a$ is at level $d_a$. Then, on the event  $\{\mathcal{T}_{a, \rm rej} = \mathcal{T'}\}$ we have
 \begin{align}
	\gamma_u \le  \frac{\alpha}{\har_{d,r}} |\mathcal{L}_a|  \tR_{\cT'}\,.
 \end{align}   
 \end{lemma}
 
 The proof of Lemma~\ref{lem:gamma-UB} follows readily from the fact that on the event $\{\mathcal{T}_{a, \rm rej} = \mathcal{T'}\}$, we have $R^{1:d}\le \tR_{\cT'}$. Also, since $u\in \cT_a$ we have $|\mathcal{L}_u| \le |\mathcal{L}_a|$. 

We next provide an upper bound for the thresholds $\alpha_u(r^*_d)$ for all nodes $u\in \cT_{a,{\rm rej}}$, which will be useful in controlling $\FSR$. For positive integer $m$, define
 \begin{align}\label{eq:tgamma}
  \tgamma_{a,m}:= \frac{\alpha}{\har_{d,r}} |\mathcal{L}_a| m\,.
 \end{align}
 Using Lemma \ref{lem:gamma-UB} and the fact $\alpha_u(r^*_d)$ is increasing in $\gamma_u$, we obtain that on the event $\{\tR_{\cT'} = m\}$, the following holds:
 \begin{align}\label{eq:talpha-2}
 \alpha_u(r^*_d)\le \talpha_{a,m}:= 
   \frac{1}{\Delta} \frac{\tgamma_{a,m}}{p(1-\frac{1}{\Delta^2})+\tgamma_{a,m} }\,.
 \end{align}
 We are now ready to upper bound the right hand side of \eqref{eq:condFSR0}.
 \begin{propo}\label{lem:partialFSRGen}
 Let $a\in \mathcal{B}^*$ and assume that the null $p$-values are
 mutually independent, and independent from the non-null
 $p$-values. 
 For our testing procedure described in Algorithm~\ref{tree algo}, the following holds true:
 \begin{align}\label{eq:partialFSRGen}
\mathbb{E}\left[\sum_{\mathcal{T'} \in S(\mathcal{T}_a)} \frac{ V_a(\mathcal{T'}) }{\tR_{\cT'}} \cdot  \prob( \mathcal{T}_{a, \rm rej} = \mathcal{T'}|\PTac)\right]\le \alpha \frac{|\cL_a|}{p}\,.
 \end{align}
 \end{propo}
 The proof of Proposition~\ref{lem:partialFSRGen} uses the
 equality~\eqref{eq:talpha} and the structural properties of the tree
 $\cT$ tree. Its proof is deferred to
 Section~\ref{proof:lem:partialFSRGen} of the appendix. The bound~\eqref{eq:partialFSR} now follows readily by applying iterative expectation to \eqref{eq:partialFSRGen}. 
	
\begin{proof}[Proof of Theorem~\ref{thm:generalFSRcontrol}]
 By using the bound \eqref{eq:partialFSR} and noting that $V = \sum_{a\in \mathcal{B}^*} V_a$, we have
	\[
	\FSR =\sum_{a\in \mathcal{B}^*}\mathbb{E}\left[\frac{V_a}{R\vee 1}\right]
	=\sum_{a\in \mathcal{B}^*}\mathbb{E}\left[\frac{V_a\, \ind(V_a>0)}{R}\right]
	\leq \sum_{a\in \mathcal{B}^*}\frac{|\mathcal{L}_a|}{p}\alpha
	=\alpha.
	\]
The result follows.
\end{proof}
\subsection{Proof of Theorem~\ref{thm:generalFSRcontrol-ext}}
Theorem~\ref{thm:generalFSRcontrol-ext} can be proved by following similar lines of the proof of Theorem \ref{thm:generalFSRcontrol} and so we omit a detailed proof here. The main difference is that in this case, the quantity $\talpha_{a,m}$ should be defined as
\begin{align}
\talpha_{a,m}:= 
   \frac{1}{\Delta}\frac{\tgamma_{a,m}}{p(1-\frac{1}{\Delta^2})+\tgamma_{a,m} } -\eps_0\,.
\end{align}
Also, the bound \eqref{eq:qUB} is updated as 
\begin{align}\label{eq:qUB-new}
q_{u,m} = \prob( u\in \mathcal{T}^m_{a, \rm rej})\le (\talpha_{a,m}+\eps_0)^{\depth(u)-\depth(a)+1} \,,
\end{align} 
and therefore similar to \eqref{eq:sum_q} we have
\begin{align}\label{eq:sum_q_new}
\sum_{u\in\cT_a} q_{u,m} &\le \frac{1}{\Delta} \cdot \frac{\Delta(\talpha_{a,m}+\eps_0) }{1-\Delta(\talpha_{a,m}+\eps_0) } = \frac{1}{\Delta p(1-\frac{1}{\Delta^2})} \tgamma_{a,m}\,,
\end{align}
which is the same bound as in \eqref{eq:sum_q}, albeit via a slightly different derivation and choice of threshold levels $\alpha_u(r)$. The rest of the proof would be identical to the proof of Theorem~\ref{thm:generalFSRcontrol}.

\subsection{Proof of Theorem~\ref{thm:reshapeFSRcontrol}}\label{proof:thm:reshapeFSRcontrol}

Let $a \in \mathcal{B}^*$, we have
\begin{align}
\mathbb{E}\left[ \frac{V_a}{R}\cdot \mathbbm{1}\{V_a > 0\}\right] & =\mathbb{E} \left[ \sum_{\Tp \in S(\mathcal{T}_a)} \frac{V_a (\Tp)}{R} \cdot \mathbbm{1}\left\{\mathcal{T}_{a, \textrm{rej}} = \Tp\right\}\right]\nonumber\\
&\leq (\Delta - \frac{1}{\Delta})\sum_{\Tp \in S(\mathcal{T}_a)} \mathbb{E} \left[ \frac{ |\Tp|}{R} \cdot \mathbbm{1}\left\{\mathcal{T}_{a, \textrm{rej}} = \Tp\right\}\right]\nonumber\\
& = (\Delta - \frac{1}{\Delta})\sum_{\Tp \in S(\mathcal{T}_a)} \sum_{u \in \Tp} \mathbb{E} \left[  \frac{\mathbbm{1}\left\{\mathcal{T}_{a, \textrm{rej}} = \Tp\right\}}{R}\right]\nonumber\\
& = (\Delta - \frac{1}{\Delta}) \sum_{u \in \mathcal{T}_a} \sum_{\Tp \in S(\mathcal{T}_a): u\in \Tp} \mathbb{E}\left[ \frac{\mathbbm{1}\left\{\mathcal{T}_{a, \textrm{rej}} = \Tp\right\}}{R}\right]\nonumber\\
& = (\Delta - \frac{1}{\Delta})  \sum_{u \in \mathcal{T}_a} \mathbb{E}\left[ \frac{\mathbbm{1} \left\{ u \in \mathcal{T}_{a, \textrm{rej}}\right\}}{R}\right]\nonumber\\
& = (\Delta - \frac{1}{\Delta}) \sum_{u \in \mathcal{T}_a} \mathbb{E}\left[ \frac{\mathbbm{1} \left\{ p_u \leq \alpha_u(r^*_d)\right\}}{R}\right]\nonumber\\
& \leq (\Delta - \frac{1}{\Delta}) \sum_{u \in \mathcal{T}_a} \mathbb{E}\left[ \frac{\mathbbm{1} \left\{ p_u \leq \alpha_u(r^*_d)\right\}}{R^{1:(d-1)}+r_d^*}\right] \nonumber\\
& = (\Delta - \frac{1}{\Delta}) \sum_{u \in \mathcal{T}_a} \mathbb{E}\left[ \frac{\mathbbm{1} \left\{ p_u \leq \frac{\alpha |\mathcal{L}_u| \beta_d(R^{1:(d-1)}+r_d^*)}{p(\Delta - \frac{1}{\Delta})(D-1)}\right\}}{R^{1:(d-1)}+r_d^*}\right],\label{eq:chain-ineq}
\end{align}
where the first inequality follows from Lemma \ref{lemma:leafbound}; the second inequality is because $R \geq R^{1:d} = R^{1:(d-1)} + r^*_d$.

Next, we will use the following proposition by Blanchard \& Roquain.

\begin{proposition}[\cite{Blanchard_2008}]\label{pro:BR}

A couple $(U,V)$ of possibly dependent nonnegative random variables such that $U$ is superuniform, i.e., $\forall t\in [0,1], \mathbb{P}(U\leq t) \leq t$, satisfy the following inequalities 
\[
\forall c>0, \mathbb{E} \left[\frac{\mathbbm{1} \left\{ U \leq c\beta(V) \right\}}{V} \right] \leq c,
\]
if $\beta(\cdot)$ is a shape function of the following form

\[
\beta(x) = \int_{0}^{x} t d\nu(t),
\]
where $\nu$ is an arbitrary probability distribution on $(0, \infty)$, and $V$ is arbitrary.

\end{proposition}

Letting $U=p_u$, $V = R^{1:(d-1)} + r^*_d$, and $c = \frac{\alpha |\mathcal{L}_u|}{p(\Delta - \frac{1}{\Delta}) (D-1)}$, we have

\begin{align*}
(\Delta - \frac{1}{\Delta}) \sum_{u \in \mathcal{T}_a} \mathbb{E}\left[ \frac{\mathbbm{1} \left\{ p_u \leq \alpha_u(r^*_d)\right\}}{R^{1:(d-1)}+r^*_d}\right]
& \leq (\Delta - \frac{1}{\Delta}) \sum_{u \in \mathcal{T}_a} \frac{\alpha |\mathcal{L}_u|}{p(\Delta - \frac{1}{\Delta})(D-1) }\\
& = \frac{\alpha}{p} \left[ \sum_{u \in \mathcal{T}_a} \frac{|\mathcal{L}_u|}{D-1} \right]\\
& \leq \frac{\alpha |\mathcal{L}_a|}{p},
\end{align*}

where the last inequality follows from 
\[
\sum_{u \in \mathcal{T}_a}|\mathcal{L}_u| \leq \sum_{d=2}^{D} \sum_{u \in \mathcal{T}^d \cap \mathcal{T}_a} |\mathcal{L}_u| = \sum_{d=2}^{D} |\mathcal{L}_a| = (D-1)|\mathcal{L}_a|.
\]

It is reasonable to use a measure $\nu$ that puts mass proportional to
$\frac{1}{x}$ only on the values that its arguments could possibly
take. By the design of the tree, we have

\[
R^{1:(d-1)} + r^*_d \geq (d-1)(\delta-1) + \delta - 1 = d(\delta-1),
\]
since at least one node has to be rejected on each depth from $1$ to $d-1$ for the algorithm to carry on to depth $d$, and 

\[ 
R^{1:(d-1)} + r^*_d \leq \sum_{u \in \mathcal{T}^{d-1}} \deg_{\mathcal{T}}(u) - 1.
\]

Therefore, 
\[
\beta_d(R^{1:(d-1)} +r^*_d) = \frac{R^{1:(d-1)} +r^*_d}{\sum_{k = d(\delta - 1)}^{\sum_{u \in \mathcal{T}^{d-1}}\deg_{\mathcal{T}}(u) -1} \frac{1}{k} }.
\]

The rest of the proof is identical to the proof of Theorem~\ref{thm:generalFSRcontrol}.

\subsubsection{Proof of Theorem \ref{thm:reshapeFSRcontrol-ext}}\label{proof:thm:reshapeFSRcontrol-ext}
The result of theorem can be derived by a similar argument used in the proof of Theorem~\ref{thm:reshapeFSRcontrol}. We leave out a detailed proof and only highlight the required modifications to the proof of Theorem~\ref{thm:reshapeFSRcontrol}.

Following the chain of inequalities in~\eqref{eq:chain-ineq} we have that for $a\in\mathcal{B}^*$,
\begin{align}
\mathbb{E}\left[ \frac{V_a}{R}\cdot \mathbbm{1}\{V_a > 0\}\right]  
& \leq (\Delta - \frac{1}{\Delta}) \sum_{u \in \mathcal{T}_a} \mathbb{E}\left[ \frac{\mathbbm{1} \left\{ p_u \leq \alpha_u(r^*_d)\right\}}{R^{1:(d-1)}+r_d^*}\right] \nonumber\\
& = (\Delta - \frac{1}{\Delta}) \sum_{u \in \mathcal{T}_a} \mathbb{E}\left[ \frac{\mathbbm{1} \left\{ p_u \leq \frac{\alpha |\mathcal{L}_u| \beta_d(R^{1:(d-1)}+r_d^*)}{p(\Delta - \frac{1}{\Delta})(D-1)} -\eps_0\right\}}{R^{1:(d-1)}+r_d^*}\right]\nonumber\\
& = (\Delta - \frac{1}{\Delta}) \sum_{u \in \mathcal{T}_a} \mathbb{E}\left[ \frac{\mathbbm{1} \left\{ p_u +\eps_0 \leq \frac{\alpha |\mathcal{L}_u| \beta_d(R^{1:(d-1)}+r_d^*)}{p(\Delta - \frac{1}{\Delta})(D-1)} -\eps_0\right\}}{R^{1:(d-1)}+r_d^*}\right]\,
\end{align}
where we used the definition of threshold function $\alpha_u(r^*_d)$ given by~\eqref{eq:threshold_reshaping_asym}. Also by theorem assumption $p_u+\eps_0$ is super-uniform because
\[
\prob(p_u+\eps_0\le t) = \prob(p_u \le t-\eps_0)\le (t-\eps_0)+\eps_0 = t\,. 
\]
Therefore we can apply Proposition \ref{pro:BR} with $U=p_u+\eps_0$, $V = R^{1:(d-1)} + r^*_d$, and $c = \frac{\alpha |\mathcal{L}_u|}{p(\Delta - \frac{1}{\Delta}) (D-1)}$.
The rest of the proof is identical to the proof of Theorem~\ref{thm:reshapeFSRcontrol} and is omitted.

\section{Proof of technical lemmas}\label{proof:lemmas}
\subsection{Proof of Lemma~\ref{lemma:FSRpowervsFDRpower}}\label{proof:lemma:FSRpowervsFDRpower}

Following the example given in Figure~\ref{fig:example_FDP}, we use solid shape for true barriers and dashed shape for achieved barriers. Therefore, $\FDP^{b}$ can be written as 

\[
\FDP^{b} = \frac{|i\in[p-1]:\; \vartheta^*_i = 0, \widehat{\vartheta}_i = 1|}{|i\in[p-1]:\; \widehat{\vartheta}_i = 1|} = \frac{\#\{\text{slots with dashed but not solid barriers}\}}{\#\{\text{slots with  dashed barriers}\}}.
\]

Similarly, we can write $\TPP^{b}$ as 

\[
\TPP^{b} = \frac{|i\in[p-1]:\; \vartheta^*_i = 1, \widehat{\vartheta}_i = 1|}{|i\in[p-1]:\; {\vartheta}^*_i = 1|}= \frac{\#\{\text{slots with solid and dashed barriers}\}}{\#\{\text{slots with solid barriers}\}},
\]
where $(\mathcal{R}^b)^{C} \coloneqq (\mathcal{H}^b_0 \cup \mathcal{H}^b_1) \setminus \mathcal{R}^b$ is the set of non-rejections.

Furthermore, we know that $n$ barriers yield $(n+1)$ groups. Hence $$\#\{\text{slots with only dashed barriers}\} = M-1$$ and $$\#\{\text{slots with only solid barriers}\} = K-1.$$ 

Fixing the solid barriers, any dashed barrier in a slot where a solid barrier does not already exist will divide one true group into two. That is to say, within each true group $C^*_i$, the number of dashed barriers, say $m$, will yield $(m+1)$ pairs of $\{i, j\}$ such that $C^*_i \cap \widehat{C}_j \neq \emptyset$. Therefore, 
\begin{align*}
\#\{\text{slots with dashed but not solid barriers}\} &= \sum_{i=1}^ K \left(\sum_{j=1}^M \mathbbm{1}\{C^*_i \cap \widehat{C}_j \neq \emptyset\}-1\right)\\
& = \sum_{i=1}^K \left(\sum_{j=1}^M \mathbbm{1}\{C^*_i \cap \widehat{C}_j \neq \emptyset\}\right)-K.
\end{align*}

Similarly, by exchanging the role of dashed and solid barriers, we also get 
\begin{align*}
\#\{\text{slots with solid but not dashed barriers}\} &= \sum_{i=1}^M \left(\sum_{j=1}^K \mathbbm{1}\{C^*_i \cap \widehat{C}_j \neq \emptyset\}-1\right) \\
&= \sum_{i=1}^M \left(\sum_{j=1}^K \mathbbm{1}\{C^*_i \cap \widehat{C}_j \neq \emptyset\}\right)-M.
\end{align*}

Finally, by plugging in the terms, we can draw the equivalence between
$\FDP^{b}$ and $\FSP$, as well as the two true positive proportions.


\subsection{Proof of Lemma \ref{lemma:FSP=V/R}}
We will prove the lemma by showing that 
\begin{equation}\label{eq:equav_of_R}
\max\left\{\sum_{u \in \mathcal{T}_{\rej}} ({\deg}_{\mathcal{T}}(u) - {\deg}_{\mathcal{T}_{\rej}}(u) )  -1, 0\right\} = M-1.
\end{equation}
and
\begin{equation}\label{eq:equav_of_V}
\sum_{u \in \mathcal{F}} \left(\deg_{\mathcal{T}}(u) -\deg_{\mathcal{T}_{\rej}}(u)\right) - \left| \mathcal{B}^*\cap \mathcal{F}\right| = \sum_{i=1}^K \left(\sum_{j=1}^M \mathbbm{1}\{C^*_i \cap \widehat{C}_j \neq \emptyset\}\right)-K,
\end{equation}

The proof is based on induction on the depth of the tree $D$.

We first prove the induction basis when $D = 2$. In this case, $\mathcal{T}$ consists in one root node $u_0$ and its children as leaves. We therefore have only one hypothesis, $\hnull{u_0}$.
\begin{itemize}
\item  If $\hnull{u_0}$ fails to be rejected, then $\mathcal{F}=\mathcal{T}_{\rej} = \emptyset$, $M=1$. Both left hand side and right hand side of equation~\eqref{eq:equav_of_R} are 0. For equation~\eqref{eq:equav_of_V}, the left hand side is clearly $0$, and the right hand side is also $0$ since $\sum_{i=1}^K \left(\sum_{j=1}^M \mathbbm{1}\{C^*_i \cap \widehat{C}_j \neq \emptyset\}\right)-K = (\sum_{i=1}^K 1) - K = 0$.
\item If $\hnull{u_0}$ is rejected, we will have $M = \deg_{\mathcal{T}}(u_0)$ and $\mathcal{T}_{\rm rej} = \{u_0\}$. Equation~\eqref{eq:equav_of_R} holds because $\sum_{u \in \mathcal{T}_{\rej}} ({\deg}_{\mathcal{T}}(u) - {\deg}_{\mathcal{T}_{\rej}}(u) )  -1 = {\deg}_{\mathcal{T}}(u_0) - {\deg}_{\mathcal{T}_{\rej}}(u_0) - 1 = M - 1$ since ${\deg}_{\mathcal{T}_{\rej}}(u_0) = 0$.  For equation~\eqref{eq:equav_of_V},  we consider two scenarios: 
\begin{itemize}
\item If $\hnull{u_0}$ is true, then $K = |\mathcal{B}^*| = 1$ and $\mathcal{F}=\mathcal{T}_{\rej} = \left\{ u_0\right\}$. So the left hand side of \eqref{eq:equav_of_V} becomes $ {\deg}_{\mathcal{T}}(u_0)  - 1 = M-1$, and the right hand side becomes $M-K = M-1$, hence the equality holds. 
\item Otherwise 
$\hnull{u_0}$ is false and $K = |\mathcal{B}^*| = \deg_{\mathcal{T}}(u_0) = M$, and $\mathcal{F}=\emptyset$. So the left hand side of \eqref{eq:equav_of_V} becomes $0$, and the right hand side becomes $M-K=0$, hence the equality holds.
\end{itemize}
\end{itemize}

Next we proceed by proving the induction step for equation \eqref{eq:equav_of_R}. Let $D>2$ be an arbitrary integer. We assume for a tree with maximum depth $\leq D-1$, identity \eqref{eq:equav_of_R} holds. We want to show that it holds for a tree with maximum depth $D$. 

Clearly, this equation holds when the root node is not rejected, i.e., $\cT_{\rej} = \emptyset$ and $M=1$. We henceforth discuss the case that the root node is rejected. In this case, equation \eqref{eq:equav_of_R} can be simplified as 

\begin{equation*}
\label{eq:equav_of_R_short} \sum_{u \in \mathcal{T}_{\rej}} ({\deg}_{\mathcal{T}}(u) - {\deg}_{\mathcal{T}_{\rej}}(u) )  = M. 
\end{equation*}

For a tree $\cT$ with maximum depth $D$, if we remove the root node, we will be left with a forest where each tree is of maximum depth less than $D$. Within each tree, we have that identity \eqref{eq:equav_of_R} holds by the induction hypothesis. We refer to the set of trees in the forest as $S_{\root}$. Furthermore, we use $M_{\cT'}, \cT' \in S_{\root}$ for the number of achieved groups in each such tree. Obviously,  

\begin{equation}\label{eq:M_sub}
\sum_{\cT' \in S_{\root}} M_{\cT'} = M.
\end{equation}

Therefore, 
\begin{align*}
&\sum_{u \in \mathcal{T}_{\rej}} ({\deg}_{\mathcal{T}}(u) - {\deg}_{\mathcal{T}_{\rej}}(u))  \\
&= \sum_{u \in \cT_{\rej}\setminus{\root}}({\deg}_{\mathcal{T}}(u) - {\deg}_{\mathcal{T}_{\rej}}(u) ) + {\deg}_{\mathcal{T}}(\root) - {\deg}_{\mathcal{T}_{\rej}}(\root) \\
& = \sum_{\cT' \in S_{\root}} \sum_{u \in \cT_{\rej} \cap \cT'}({\deg}_{\mathcal{T}}(u) - {\deg}_{\mathcal{T}_{\rej}}(u) ) + \deg_{\cT}(\root) - \deg_{\cT_{\rej}}(\root) \\
& = \sum_{\substack{\cT' \in S_{\root}\\ \cT' \cap \cT_{\rej} \neq \emptyset}} \sum_{u \in \cT_{\rej} \cap \cT'}({\deg}_{\mathcal{T}}(u) - {\deg}_{\mathcal{T}_{\rej}}(u) ) + \deg_{\cT}(\root) - \deg_{\cT_{\rej}}(\root) \\
& = \sum_{\substack{\cT' \in S_{\root}\\ \cT' \cap \cT_{\rej} \neq \emptyset}}M_{\cT'}  + \deg_{\cT}(\root) - \deg_{\cT_{\rej}}(\root) \\
& = \sum_{\substack{\cT' \in S_{\root}\\ \cT' \cap \cT_{\rej} \neq \emptyset}}M_{\cT'} + \sum_{\substack{\cT' \in S_{\root}\\ \cT' \cap \cT_{\rej} = \emptyset}}M_{\cT'}  \\
& = M,
\end{align*}
where the fourth equality is by the induction hypothesis; the fifth
equality holds because there are $\deg_{\cT}(\root) -
\deg_{\cT_{\rej}}(\root)$ subtrees $\cT' \in S_{\root}$ such that $\cT' \cap \cT_{\rej} = \emptyset$, and their $M_{\cT'} = 1$; the last equality follows from \eqref{eq:M_sub}. This proves the induction step and hence completes the proof of identity \eqref{eq:equav_of_R}.

We next proceed to prove \eqref{eq:equav_of_V}. Suppose that the
induction hypothesis holds for trees with depth at most $D-1$. We want
to prove it for trees of depth $D$. Note that this identity trivially
holds when the root is not rejected, and therefore we focus on the
case where the root is rejected. There are two scenarios: (1) the root
is a true rejection, or (2) the root is a false rejection.

We first assume the root is a true rejection. Then we have 

\begin{equation}\label{eq:K_sub}
K = \sum_{\cT' \in S_{\root}} K_{\cT'},
\end{equation}

where $K_{\cT'} \geq 1$ is defined as the number of true groups in each $\cT' \in S_{\root}$.

Then the left hand side of \eqref{eq:equav_of_V} becomes
\begin{align*}
&\sum_{u \in \mathcal{F}} \left(\deg_{\mathcal{T}}(u) -\deg_{\mathcal{T}_{\rej}}(u)\right) - \left| \mathcal{B}^*\cap \mathcal{F}\right|\\ &= \sum_{\cT' \in S_{\root}} \left(\sum_{u \in \mathcal{F}\cap \cT'} \deg_{\mathcal{T}}(u) -\deg_{\mathcal{T}_{\rej}}(u) - | \mathcal{B}^* \cap \mathcal{F} \cap \cT'| \right)\\
& = \sum_{\cT' \in S_{\root}} \left[ \sum_{1\leq i \leq K_{\cT'}} \left(\sum_{1\leq j \leq M_{\cT'} }\mathbbm{1}\{C^*_i \cap \widehat{C}_j \neq \emptyset\}\right) - K_{\cT'}\right]\\
& = \sum_{1\leq i \leq K} \left(\sum_{1\leq j \leq M }\mathbbm{1}\{C^*_i \cap \widehat{C}_j \neq \emptyset\}\right) - K,
\end{align*}
where the first equality holds because $\cT' \in S_{\root}$ are
disjoint from each other and the root is not in $\mathcal B^*\cap
\mathcal F$; the second equality follows from the induction hypothesis, and the last equality follows from \eqref{eq:M_sub} and \eqref{eq:K_sub}.

For the case where the root is a false rejection, we have $K=1$, $\mathcal{B}^* = \{\root\}$ and any rejection is a false rejection ($\mathcal{F} = \cT_{\rm rej}$). We write
\begin{align*}
\sum_{u \in \mathcal{F}} \left(\deg_{\mathcal{T}}(u) -\deg_{\mathcal{T}_{\rej}}(u)\right) - \left| \mathcal{B}^*\cap \mathcal{F}\right|= \sum_{u \in \cT_{\rm rej}} \left(\deg_{\mathcal{T}}(u) -\deg_{\mathcal{T}_{\rej}}(u)\right) - 1  = M-1\,,
\end{align*}
where in the last step we used identity~\eqref{eq:equav_of_R}. On the other hand, in this case there is only one true group ($K=1$) which consists of all leaves. Therefore, any returned group $\widehat{C}_j$ will intersect with it and we get
\begin{align*}
\sum_{i=1}^K \left(\sum_{j=1}^M \mathbbm{1}\{C^*_i \cap \widehat{C}_j \neq \emptyset\}\right) - 1 = M-1\,.
\end{align*}
Comparing the previous two equations implies that identity~\eqref{eq:equav_of_V} holds for the tree $\cT$. This completes the induction step and hence proves identity~\eqref{eq:equav_of_V}.

\subsection{Proof of Lemma \ref{coro:binaryFSP=V/R}}
The proof of Lemma~\ref{coro:binaryFSP=V/R} follows from Lemma~\ref{lemma:FSP=V/R} and that $\deg_{\mathcal{T}}(u) = 2$, for all non-leaf nodes $u\in \mathcal{T}$. It suffices to show 

\begin{align}\label{eq:V_binary_simple}
\sum_{u \in \mathcal{F}} \left(\deg_{\mathcal{T}}(u) -\deg_{\mathcal{T}_{\rej}}(u)\right) - \left| \mathcal{B}^*\cap \mathcal{F}\right| = |\mathcal{F}|,
\end{align}

and 

\begin{align}\label{eq:R_binary_simple}
\max\left\{\sum_{u \in \mathcal{T}_{\rej}} ({\deg}_{\mathcal{T}}(u) - {\deg}_{\mathcal{T}_{\rej}}(u) )  -1, 0\right\}  = |\mathcal{T}_{\textrm{rej}}|.
\end{align}

To prove equation~\eqref{eq:V_binary_simple}, note that if a node is falsely rejected all of its rejected children are also false rejections. Therefore, $\sum_{u\in \mathcal{F}} \deg_{\cT_{\rm rej}}(u)$ counts the total number of edges where both nodes of it are in $\mathcal{F}$. Hence, 
\begin{align*}
&\sum_{u \in \mathcal{F}} \left(\deg_{\mathcal{T}}(u) -\deg_{\mathcal{T}_{\rej}}(u)\right) - \left| \mathcal{B}^*\cap \mathcal{F}\right|\\
& = 2|\mathcal{F}| - |\left\{u: u\in \mathcal{F}, \parent(u)\in \mathcal{F} \right\}| - \left| \mathcal{B}^*\cap \mathcal{F}\right|\\
& = 2|\mathcal{F}| - |\left\{u: u\in \mathcal{F} , \parent(u)\in \mathcal{F}\right\}| - |\left\{u: u\in \mathcal{F}, \parent(u)\notin \mathcal{F}\right\}|\\
& = 2|\mathcal{F}| - |\mathcal{F}| \\
& = |\mathcal{F}|.
\end{align*}

Equation~\eqref{eq:R_binary_simple} holds trivially when $|\cT_{\rm{rej}}|=0$. When $|\cT_{\textrm{rej}}|>0$, the root node is rejected, and we write
\begin{align*}
&\sum_{u \in \mathcal{T}_{\rej}} ({\deg}_{\mathcal{T}}(u) - {\deg}_{\mathcal{T}_{\rej}}(u) )  -1 \\
& = 2|\mathcal{T}_{\textrm{rej}}|  - \sum_{u \in \mathcal{T}_{\rej}} {\deg}_{\mathcal{T}_{\rej}}(u) -1\\
& = 2|\mathcal{T}_{\textrm{rej}}| - (|\mathcal{T}_{\textrm{rej}}|-1)-1\\
& = |\mathcal{T}_{\textrm{rej}}|.
\end{align*}
This completes the proof.
\subsection{Proof of Lemma \ref{lemma:ANOVA}}
We use the shorthand $\ell_u:= |\cL_u|$ for a node $u$. 
Define the random vector $\boldsymbol{w}\in\real^{\Delta_a}$ with elements
$w_u=\ell_u^{1/2}\bar y_u$ and the fixed unit vector
$\boldsymbol{r}\in\real^{\Delta_a}$ with elements $r_u=(\ell_u/\ell_a)^{1/2}$.
We have
\begin{equation}
\boldsymbol{r}^{\top} \boldsymbol{w}=\sum_{u\in\child(a)}(\ell_u/\ell_a)^{1/2}(\ell_u^{\frac{1}{2}}\bar
y_u)=\ell_a^{-1/2}\sum_{u\in\child(a)}\ell_u\bar y_u=\ell_a^{1/2}\bar
y_a,\label{eq:rW}
\end{equation}
from which it follows that
$$
\sum_{u\in\child(a)}\ell_u(\bar y_u - \bar y_a)^2 = \sum_{u\in\child(a)} (\ell_u^{1/2}(\bar y_u - \bar y_a))^2
= \sum_{u\in\child(a)} (w_u - r_u \boldsymbol{r}^{\top} \boldsymbol{w})^2 
=\|(\boldsymbol{I}_{\Delta_a}-\boldsymbol{r}\boldsymbol{r}^{\top})\boldsymbol{w}\|^2.
$$
The random vector $\boldsymbol{w}$ is multivariate normal with
$\E[w_u]=\ell_u^{1/2}\bar\theta_u$ , where $\bar\theta_u = \frac{1}{|\cL_u|}\sum_{i\in\cL_u} \theta_i$ is the average of parameters on the leave nodes $\cL_u$. In addition, $\cov(\boldsymbol{w})=\sigma^2
\boldsymbol{I}_{\Delta_a}$.  Taking the expectation of \eqref{eq:rW} establishes
that 
\[
\E[(\boldsymbol{r}\boldsymbol{r}^{\top} \boldsymbol{w})_u]= \E[ r_u (\boldsymbol{r}^{\top} \boldsymbol{w})] = (\ell_u/\ell_a)^{1/2}\ell_a^{1/2}\E[\bar y_a]
= \ell_u^{1/2} \bar\theta_a.
\] 
Under $\H_a$, we have $\bar\theta_u = \bar\theta_a$ and thus
$$
(\boldsymbol{I}_{\Delta_a}-\boldsymbol{r}\boldsymbol{r}^{\top})\boldsymbol{w}\sim \normal\left(\boldsymbol{0},\sigma^2(\boldsymbol{I}_{\Delta_a}-\boldsymbol{r}\boldsymbol{r}^{\top})\right)\,,
$$
where we use the fact that $\|\boldsymbol{r}\|_2 = 1$ and so $\boldsymbol{I}_{\Delta_a}-\boldsymbol{r}\boldsymbol{r}^{\top}$ is a projection matrix.
This establishes that 
$$
\sum_{u\in\child(a)}\ell_u(\bar y_u - \bar y_a)^2\sim \sigma^2\chi^2_{\Delta_a-1}
$$
under $\H_a$, meaning that $p_a$ is uniform.  Now consider some node
$b\neq a$.  If $\L_a\cap\L_b=\emptyset$, then $p_a$ and $p_b$ are
clearly independent (because they depend only on $\boldsymbol{y}_{\L_a}$ and
$\boldsymbol{y}_{\L_b}$, respectively).  Thus, it remains to consider the case that
$\L_a\subset \L_b$ (i.e., $a$ is a descendant of $b$).  There must exist
$v\in\child(b)$ with $\L_a\subseteq \L_v\subset \L_b$.  From
\eqref{eq:anova-pvalue}, $p_b=f(\bar y_v, \boldsymbol{y}_{\L_b\setminus\L_v})$.  Since
$(\L_b\setminus \L_v)\cap \L_a=\emptyset$, we know that $p_a$ is
independent of $\boldsymbol{y}_{\L_b\setminus\L_v}$.  It therefore remains to show
that $p_a$ is also independent of $\bar y_v$.  To do so, observe that
\begin{equation}
\ell_v\bar y_v = \sum_{i\in\L_a}y_i +
\sum_{i\in\L_v\setminus\L_a}y_i=\ell_a^{1/2}\boldsymbol{r}^{\top}\boldsymbol{w} +
\sum_{i\in\L_v\setminus\L_a}y_i.\label{eq:Yv}
\end{equation}
Thus,
\begin{align*}
\cov\left([\boldsymbol{I}_{\Delta_a}-\boldsymbol{r}\boldsymbol{r}^T]\boldsymbol{w},\bar
  y_v\right)&=\ell_v^{-1}\cov\left([\boldsymbol{I}_{\Delta_a}-\boldsymbol{r}\boldsymbol{r}^{\top}]\boldsymbol{w},
              \ell_v\bar y_v\right)\\
  &=\ell_a^{1/2}\ell_v^{-1}\cov\left([\boldsymbol{I}_{\Delta_a}-\boldsymbol{r}\boldsymbol{r}^{\top}]\boldsymbol{w},
              \boldsymbol{r}^{\top}\boldsymbol{w}\right)\\
  &=\sigma^2\ell_a^{1/2}\ell_v^{-1}[\boldsymbol{I}_{\Delta_a}-\boldsymbol{r}\boldsymbol{r}^{\top}]\boldsymbol{r}\\
  &=0,
\end{align*}
where the first equality follows from observing that the second term
in \eqref{eq:Yv} is independent of $\boldsymbol{w}$ (which depends only on
$\boldsymbol{y}_{\L_a}$) and the second inequality uses that
$\cov(\boldsymbol{w})=\sigma^2 \boldsymbol{I}_{\Delta_a}$.  This establishes that $p_a$ is
independent of $p_b$.

\subsection{Proof of Proposition~\ref{propo:p-val}}\label{proof:propo:p-val}
Note that for any node $u$, we have $\|\bG_u\|_2 = 1$ since $\bG_u$ is a projection matrix.
Also, by using~\cite[Lemma 2]{guo2019group} (which itself follows from~\cite[Lemma 1]{cai2019individualized}), we have
\begin{align}
\|\bG_u \widehat{\btheta}\|_2 \le c_0 (\widehat{\boldsymbol{b}}^{\top} \widehat{\boldsymbol{\Sigma}} \widehat{\boldsymbol{b}})^{1/2}\,,
\end{align} 
for some constant $c_0>0$.  This inequality follows by analyzing the optimization~\eqref{eq:ConvexOptimization} which is used to define the direction $ \widehat{\boldsymbol{b}}$. Therefore, for any node $u$, we obtain
\begin{align}
\frac{\|\bG_u\widehat{\btheta}\|_2+\|\bG_u\|_2 }{\sqrt{ \widehat{\var}_{\tau}(\widehat{Q}^{\rm d}_u) }} 
&\le\frac{ c_0 (\widehat{\boldsymbol{b}}^{\top} \widehat{\boldsymbol{\Sigma}} \widehat{\boldsymbol{b}})^{1/2} + 1}{\sqrt{ \widehat{\var}_{\tau}(\widehat{Q}^{\rm d}_u) }} \nonumber\\
&= \frac{ c_0 (\widehat{\boldsymbol{b}}^{\top} \widehat{\boldsymbol{\Sigma}} \widehat{\boldsymbol{b}})^{1/2} + 1}{\sqrt{\frac{4\widehat{\sigma}^2}{n}  \widehat{\boldsymbol{b}}^{\top} \widehat{\boldsymbol{\Sigma}} \widehat{\boldsymbol{b}} + \frac{\tau}{n} }} \nonumber\\
&\le \frac{c_0\sqrt{n}}{2\widehat{\sigma}} + \sqrt{\frac{n}{\tau}}\,.\label{eq:dummy-G}
\end{align}

Define $x:=\Phi^{-1}(1-\frac{t}{2})$ and 
\begin{align}\label{eq:etan}
\eta_n := c_1 \left(\frac{c_0}{2\widehat{\sigma}} + \sqrt{\frac{1}{\tau}}\right) \frac{s_0\log p}{\sqrt{n}} \,,
\end{align}
with $c_1$ given in~\eqref{eq:bias}. Under the null hypothesis $\widetilde{\mathcal{H}}_{0,u}$ (or equivalently ${\mathcal{H}}_{0,u}$), we have for all $t\in [0,1]$,
\begin{align}
\prob(p_u\le t) &= \prob\left(\Phi^{-1}(1-\frac{t}{2}) \le \frac{|\widehat{Q}^{\rm d}_u|}{\sqrt{\widehat{\var}_\tau(\widehat{Q}_u) }}\right)\nonumber\\
&= \prob\left(x\le \frac{|\widehat{Q}^{\rm d}_u|}{\sqrt{\widehat{\var}_\tau(\widehat{Q}_u) }}\right)\nonumber\\
&= \prob\left(x\le \frac{|Z_u+\Delta_u|}{\sqrt{\widehat{\var}_\tau(\widehat{Q}_u) }}\right)\nonumber\\
&\le \prob\left(x -\eta_n\le \frac{|Z_u|}{\sqrt{\widehat{\var}_\tau(\widehat{Q}_u) }}\right) + \prob\left(\eta_n \le \frac{|\Delta_u|}{\sqrt{\widehat{\var}_\tau(\widehat{Q}_u) }}\right).\label{eq:p-unifB}
\end{align}
By using the bias bound~\eqref{eq:bias}, together with \eqref{eq:dummy-G} and definition of $\eta_n$ given by~\eqref{eq:etan}, we have
\begin{align}\label{eq:biasB}
 \prob\left(\eta_n \le \frac{|\Delta_u|}{\sqrt{\widehat{\var}_\tau(\widehat{Q}_u) }}\right) \le 2pe^{-c_2n}\,,
\end{align}
for all nodes $u$. In addition, 
\begin{align}
&\prob\left(x -\eta_n\le \frac{|Z_u|}{\sqrt{\widehat{\var}_\tau(\widehat{Q}_u) }}\right)\nonumber\\
&\le \prob\left(x -\eta_n\le \  \frac{|Z_u|}{\sqrt{\frac{4\widehat{\sigma}^2}{n}  \widehat{\boldsymbol{b}}^{\top} \widehat{\boldsymbol{\Sigma}} \widehat{\boldsymbol{b}} }}\right) = \prob\left(x -\eta_n\le \  \frac{\sigma |Z_u|}{\widehat{\sigma}\sqrt{\var(\widehat{Q}^{\rm d}_u)}}\right)\nonumber\\
&\le \prob\left((x -\eta_n)(1-\eps)\le \  \frac{|Z_u|}{\sqrt{\var(\widehat{Q}^{\rm d}_u)} }\right) + \prob\left(\Big|\frac{\widehat{\sigma}}{\sigma} - 1\Big| \ge \eps \right)\nonumber\\
&= 2\Phi(\eps x - x +\eta_n - \eps \eta_n) + \prob\left(\Big|\frac{\widehat{\sigma}}{\sigma} - 1\Big| \ge \eps \right)\label{eq:ZB}
\end{align}
Combining~\eqref{eq:p-unifB}, \eqref{eq:biasB} and \eqref{eq:ZB} we obtain
\[
\prob(p_u\le t) \le 2\Phi(\eps x - x +\eta_n - \eps \eta_n) + \prob\left(\Big|\frac{\widehat{\sigma}}{\sigma} - 1\Big| \ge \eps \right) + 2pe^{-c_2n}\,.
\]
Note that the right-hand side of the above equation does not depend on the node $u$. In other words, it is a uniform bound for all nodes. Under the condition $s_0(\log p)/\sqrt{n}\to 0$, we have $\eta_n\to 0$ as $n\to \infty$. Therefore, for any fixed $\eps_0>0$, by choosing $\eps>0$ small enough and $n_0 = n_0(\eps)$ large enough we can ensure that for all $n\ge n_0$
\[
\prob(p_u\le t)\le 2\Phi(-x) + \eps_0 = 2(1-\Phi(x)) + \eps_0 = t+ \eps_0\,,
\]
for all nodes $u$.
	
\section{Proof of Proposition \ref{lem:partialFSRGen}}\label{proof:lem:partialFSRGen}
For depth $d$ we define the quantities $L_d:= R^{1:(d-1)}+r^*_d$ and $U_d:= p-1-\left(\sum_{u\in\cT^d} \deg_\cT(u) - |\cT^d| - r^*_d\right)$. For node $a\in \mathcal{B}^*$ with $\depth(a) = d$, we write
 \begin{align}
&\sum_{\mathcal{T'} \in S(\mathcal{T}_a)} \frac{ V_a(\mathcal{T'}) }{\tR_{\cT'}} \cdot  \prob( \mathcal{T}_{a, {\rm rej}} =\cT'|\PTac) \nonumber\\
&\stackrel{(a)}{\le} (\Delta-\frac{1}{\Delta}) \sum_{\mathcal{T'} \in S(\mathcal{T}_a)}  \frac{|\cT'|}{\tR_{\cT'}}\, \prob( \mathcal{T}_{a, {\rm rej}} =\cT'|\PTac)\nonumber\\
&= (\Delta-\frac{1}{\Delta}) \sum_{\mathcal{T'} \in S(\mathcal{T}_a)} \sum_{u\in\cT'}\frac{1}{\tR_{\cT'}}\, \prob( \mathcal{T}_{a, {\rm rej}} =\cT'|\PTac)\nonumber\\
&= (\Delta-\frac{1}{\Delta}) \sum_{u\in\cT_a} \sum_{\mathcal{T'} \in S(\mathcal{T}_a):u\in\cT'} \frac{1}{\tR_{\cT'}}\, \prob( \mathcal{T}_{a, {\rm rej}} =\cT'|\PTac)\nonumber\\
&\stackrel{(b)}{\le} (\Delta-\frac{1}{\Delta}) \sum_{u\in\cT_a}\; \sum_{m=L_d}^{U_d}\; \sum_{\mathcal{T'} \in S(\mathcal{T}_a): u\in \cT'} \frac{1}{\tR_{\cT'}}\, \prob( \mathcal{T}_{a, {\rm rej}} =\cT'|\PTac)\ind(\tR_{\cT'} = m)\nonumber\\
&= (\Delta-\frac{1}{\Delta})  \sum_{u\in\cT_a} \; \sum_{m=L_d}^{U_d}\;  \frac{1}{m}\sum_{\mathcal{T'} \in S(\mathcal{T}_a)} \prob( \mathcal{T}_{a, {\rm rej}} =\cT'|\PTac)\ind(\tR_{\cT'} = m, u\in \cT')\nonumber\\
&= (\Delta-\frac{1}{\Delta})  \sum_{u\in\cT_a}  \sum_{m=L_d}^{U_d}  \frac{1}{m} \prob( \tR_{\mathcal{T}_{a, {\rm rej}}} = m, u\in \mathcal{T}_{a, {\rm rej}}|\PTac)\,.\label{eq:chain1}
 \end{align}
Here $(a)$ follows from Lemma~\ref{lemma:rejectionnumberbound}, and
$(b)$ holds since for $\cT'\in S(\cT_a)$, 
the number of total rejections $\tR_{\cT'}$ satisfies
\[
L_d  \le \tR_{\cT'} \le U_d\,.
\]
The lower bound holds trivially since  $\cT'\in S(\cT_a)$ and $\depth(a) =d$.
The number of splits made by algorithm up to level $d$ is $R^{1:d} = R^{1:(d-1)} + r^*_d$ by using Equation~\eqref{eq:R1d}.
For the upper bound, note that one can split the $p$ leaves
at most $p-1$ times. Now focusing on nodes in depth $d$, rejecting a node $u$ results in $\deg_\cT(u) - 1$ additional splits.
So the nodes in depth $d$ can make up to $\sum_{u\in\cT^d} \deg_\cT(u) - |\cT^d|$ additional splits, while the algorithm makes $r^*_d$ additional splits as we discussed  in Equation~\eqref{eq:R1d}. So the difference between these two quantities, $\sum_{u\in\cT^d} \deg_\cT(u) - |\cT^d| - r^*_d$, is the number of potential splits that the testing rule has missed while testing nodes at depth $d$. This argument implies that the total number of splits can go up to $U_d = p-1 - (\sum_{u\in\cT^d} \deg_\cT(u) - |\cT^d| - r^*_d).$ 

Now by using bound \eqref{eq:talpha-2}, on the event $\{\tR_{\cT'} =
m\}$ we have $\alpha_u(r^*_d) \le \talpha_{a,m}$. Define
$\mathcal{T}^m_{a, \rm rej} $ as the rejection subtree as if the test
levels $\alpha_u(r^*_d)$ are replaced by $\talpha_{a,m}$. Therefore
$\mathcal{T}_{a, \rm rej} \subseteq \mathcal{T}^m_{a, \rm rej}$, which
implies
$$
\prob( \tR_{\mathcal{T}_{a, {\rm rej}}} = m, u\in \mathcal{T}_{a, {\rm
    rej}}|\PTac)\le \prob( \tR_{\mathcal{T}_{a, {\rm rej}}} = m, u\in \mathcal{T}^m_{a, {\rm rej}}|\PTac).
$$
Combining this inequality with \eqref{eq:chain1} and taking the
expectation gives
\begin{align}
\mathbb E\left[\sum_{\mathcal{T'} \in S(\mathcal{T}_a)} \frac{
    V_a(\mathcal{T'}) }{\tR_{\cT'}} \cdot  \prob( \mathcal{T}_{a, {\rm
      rej}} =\cT'|\PTac)\right]\le  (\Delta-\frac{1}{\Delta})
  \sum_{u\in\cT_a}  \sum_{m=L_d}^{U_d}  \frac{1}{m}
  \prob( \tR_{\mathcal{T}_{a, {\rm rej}}} = m, u\in \mathcal{T}^m_{a,
  {\rm rej}}).\label{eq:expectation}
\end{align}

Focusing on the innermost summation, we have
\begin{align}
& \sum_{m=L_d }^{U_d}  \frac{1}{m} \prob( \tR_{\mathcal{T}_{a, {\rm rej}}} = m, u\in \mathcal{T}^m_{a, {\rm rej}})\nonumber\\
& = \sum_{m=L_d}^{U_d}  \frac{1}{m} \left[\prob( \tR_{\mathcal{T}_{a, {\rm rej}}} \ge m, u\in \mathcal{T}^m_{a, {\rm rej}}) - \prob( \tR_{\mathcal{T}_{a, {\rm rej}}} \ge m+1, u\in \mathcal{T}^m_{a, {\rm rej}})\right]\nonumber\\
&= \sum_{m=L_d }^{U_d}  \frac{1}{m} \prob( \tR_{\mathcal{T}_{a, {\rm rej}}} \ge m, u\in \mathcal{T}^m_{a, {\rm rej}})
- \sum_{m'=L_d+1  }^{U_d+1}  \frac{1}{m'-1} \prob( \tR_{\mathcal{T}_{a, {\rm rej}}} \ge m', u\in \mathcal{T}^{m'-1}_{a, {\rm rej}})\nonumber\\
 &= \sum_{m=L_d +1}^{U_d}   \left[ \frac{1}{m} \prob( \tR_{\mathcal{T}_{a, {\rm rej}}} \ge m, u\in \mathcal{T}^m_{a, {\rm rej}}) - \frac{1}{m-1} \prob( \tR_{\mathcal{T}_{a, {\rm rej}}} \ge m, u\in \mathcal{T}^{m-1}_{a, {\rm rej}})\right]\nonumber\\
 &\quad +\frac{1}{L_d } \prob( \tR_{\mathcal{T}_{a, {\rm rej}}} \ge L_d  , u\in \mathcal{T}^{L_d  }_{a, {\rm rej}})
 -\frac{1}{U_d}  \prob( \tR_{\mathcal{T}_{a, {\rm rej}}} \ge U_d+1, u\in \mathcal{T}^{U_d}_{a, {\rm rej}})\nonumber\\
 &\le \sum_{m=L_d  +1}^{U_d}   \left[ \frac{1}{m} \prob( \tR_{\mathcal{T}_{a, {\rm rej}}} \ge m, u\in \mathcal{T}^m_{a, {\rm rej}}) - \frac{1}{m-1} \prob( \tR_{\mathcal{T}_{a, {\rm rej}}} \ge m, u\in \mathcal{T}^{m-1}_{a, {\rm rej}})\right] \nonumber\\
 &\quad+\frac{1}{L_d } \prob( u\in \mathcal{T}^{L_d }_{a, {\rm rej}})\nonumber\\
 &\le \sum_{m=L_d  +1}^{U_d}    \frac{1}{m} \prob\left( \tR_{\mathcal{T}_{a, {\rm rej}}} \ge m, u\in \mathcal{T}^m_{a, {\rm rej}}\backslash\mathcal{T}^{m-1}_{a, {\rm rej}}\right)+\frac{1}{L_d} \prob( u\in \mathcal{T}^{L_d  }_{a, {\rm rej}})\nonumber\\
 &\le \sum_{m=L_d  +1}^{U_d}  \frac{1}{m} \prob\left( u\in \mathcal{T}^m_{a, {\rm rej}}\backslash\mathcal{T}^{m-1}_{a, {\rm rej}}\right) +\frac{1}{L_d} \prob( u\in \mathcal{T}^{L_d}_{a, {\rm rej}})\,,\label{eq:chain2}
\end{align} 
where in the last equality we used the observation $\cT_{a,{\rm rej}}^{m-1}\subseteq \cT_{a,{\rm rej}}^{m}$, since $\alpha_{u,m}$ is increasing in $m$. 

For exposition purposes, we define the shorthand $q_{u,m} = \prob( u\in \mathcal{T}^m_{a, \rm rej})$ for $u\in \mathcal{T}_a$ and $m\ge 1$. Then, from the chain of inequalities in~\eqref{eq:chain2} we get
\begin{align}\label{eq:chain3}
 & \sum_{m=L_d }^{U_d}  \frac{1}{m} \prob( \tR_{\mathcal{T}_{a, {\rm rej}}} = m, u\in \mathcal{T}^m_{a, {\rm rej}}) \nonumber\\
 &\le \sum_{m=L_d  +1}^{U_d}  \frac{1}{m} \prob\left( u\in \mathcal{T}^m_{a, {\rm rej}}\backslash\mathcal{T}^{m-1}_{a, {\rm rej}}\right) +\frac{1}{L_d} \prob( u\in \mathcal{T}^{L_d }_{a, {\rm rej}})\nonumber\\
  &\le \sum_{m=L_d  +1}^{U_d}  \frac{1}{m} (q_{u,m}-q_{u,m-1}) +\frac{1}{L_d } q_{u, L_d }\nonumber\\
    &= \sum_{m=L_d }^{U_d}  \frac{1}{m} q_{u,m} - \sum_{m=L_d  }^{U_d-1} \frac{1}{m+1} q_{u,m}\nonumber\\
       &= \frac{1}{U_d}q_{u,U_d} + \sum_{m=L_d }^{U_d-1}  \left(\frac{1}{m}-\frac{1}{m+1}\right) q_{u,m} .
\end{align}

By deploying~\eqref{eq:chain3} in the bound~\eqref{eq:expectation}, we get
\begin{align}
&  \mathbb E\left[\sum_{\mathcal{T'} \in S(\mathcal{T}_a)} \frac{
    V_a(\mathcal{T'}) }{\tR_{\cT'}} \cdot  \prob( \mathcal{T}_{a, {\rm
      rej}} =\cT'|\PTac)\right] \nonumber\\
&\le (\Delta-\frac{1}{\Delta})  \sum_{u\in\cT_a} \left( \frac{1}{U_d}q_{u,U_d} + \sum_{m=L_d }^{U_d-1}  \left(\frac{1}{m}-\frac{1}{m+1}\right) q_{u,m}\right) \nonumber\\ 
&= (\Delta-\frac{1}{\Delta})  \left( \frac{1}{U_d}  \sum_{u\in\cT_a} q_{u,U_d} + \sum_{m=L_d  }^{U_d-1}  \frac{1}{m(m+1)}  \sum_{u\in\cT_a} q_{u,m}\right) \,.\label{eq:chain4}
\end{align}

Our next step is to upper bound $\sum_{u\in\cT_a} q_{u,m}$ which is the subject of the following lemma.
\begin{lemma}\label{lem:sum-pu}
For any integer $m\ge1$ we have
\[
\sum_{u\in\cT_a} q_{u,m} \le  \frac{\tgamma_{a,m}}{p(\Delta-\frac{1}{\Delta})}\,,
\]
where $\tgamma_{a,m}$ is given by~\eqref{eq:tgamma}.
\end{lemma}
The proof of Lemma~\ref{lem:sum-pu} is deferred to Section~\ref{proof:lem:sum-pu}.

By virtue of Lemma~\ref{lem:sum-pu} and \eqref{eq:chain4}, we have
\begin{align}
\mathbb E\left[\sum_{\mathcal{T'} \in S(\mathcal{T}_a)} \frac{
    V_a(\mathcal{T'}) }{\tR_{\cT'}} \cdot  \prob( \mathcal{T}_{a, {\rm
      rej}} =\cT'|\PTac)\right]
&\le\frac{1}{p}\left(\frac{1}{U_d} \tgamma_{a,U_d} +  \sum_{m=L_d }^{U_d-1} \frac{1}{m(m+1)}  \tgamma_{a,m} \right)
\nonumber\\
&=\frac{\alpha |\cL_a|}{p \har_{d,r} }\left(1+  \sum_{m=L_d  }^{U_d-1} \frac{1}{m+1}  \right)\nonumber\\
&=\frac{\alpha |\cL_a|}{p \har_{d,r} }\left(1+  \sum_{m=L_d+1}^{U_d} \frac{1}{m}  \right)\nonumber\\
&= \frac{\alpha |\cL_a|}{p }\,.
\end{align}

\subsection{Proof of Lemma~\ref{lem:sum-pu}}\label{proof:lem:sum-pu}
Since $a\in \mathcal{B}^*$, any node $u\in \cT_a$ is a true null and hence it has a super uniform $p$-value, i.e. for any $x\in [0,1]$ we have $\prob(p_u\le x)\le x$. In addition, by our assumption the null $p$-values are independent and if a node $u$ is rejected so are the nodes on the path from node $a$ to it.
Therefore,
\begin{align}\label{eq:qUB}
q_{u,m} = \prob( u\in \mathcal{T}^m_{a, \rm rej}) \le \talpha_{a,m}^{\depth(u)-\depth(a)+1}  \,.
\end{align}  
Here we used the fact that the rejection thresholds in $\mathcal{T}^m_{a, \rm rej}$ are set to $\talpha_{a,m}$. 

Also, since the node degrees  in $\cT$ are at most $\Delta$, the number of nodes in subtree $\cT_a$ that are depth $d$ of the tree $\cT$ is at most $\Delta^{d-\depth(a)}$. We therefore have
\begin{align}\label{eq:sum_q}
\sum_{u\in\cT_a} q_{u,m} &\le   \sum_{d= \depth(a)}^{D}  \Delta^{d-\depth(a)} \talpha_{a,m}^{d-\depth(a)+1}\nonumber\\
&\le  \sum_{d=1}^{\infty}  \Delta^{d-1} \talpha_{a,m}^{d} \nonumber\\
& = \frac{1}{\Delta} \frac{\Delta\talpha_{a,m} }{1-\Delta\talpha_{a,m} } \nonumber\\
&= \frac{\tgamma_{a,m}}{p(\Delta-\frac{1}{\Delta})}\,,
\end{align}
which completes the proof.

\section{Some useful lemmas}\label{section:lemma_proof}

\begin{lemma}\label{lemma:leafbound}
	Consider a tree $\mathcal{T}$ with maximum degree $\Delta$. 
	Denote by $\cL$ the set of leaf nodes in  $\cT$. We then have
	\[
	|\cL| \leq \frac{(\Delta -1) |\cT|+1}{\Delta},
	\]
	where $|\cT|$ denotes the number of nodes in $\cT$.
\end{lemma}

\begin{proof} Recall that the degree of a node $u$ is the number of its children in the tree. The leaves are of zero degree and the other nodes are of maximum degree $\Delta$. Therefore,
	
	\[
	(|\mathcal{T}|-p)\Delta \geq \sum_{u \in \mathcal{T}} \deg_{\mathcal{T}}(u) = |\mathcal{T}|-1 \,.	
	\]
	By rearranging the terms we get
	\[
	p \leq \frac{(\Delta - 1)\cdot |\mathcal{T}| + 1}{\Delta}.
	\]
	
\end{proof}

\begin{lemma}\label{lemma:rejectionnumberbound}
	Consider a tree $\mathcal{T}$ with maximum degree $\Delta$.  For $\Tp$, a subtree of $\mathcal{T}$, define
	\[
	V(\Tp)=\sum_{u \in \Tp} \left(\deg_{\mathcal{T}}(u) - \deg_{\Tp}(u)\right) - 1\,.
	\]
We then have the following bound on $V(\Tp)$:
	\[
	V(\mathcal{T'}) \leq \frac{(\Delta^2 - 1)\cdot |\Tp| + 1}{\Delta} - 1 \leq \left(\Delta-\frac{1}{\Delta}\right) |\Tp|\,,
	\]
	where $|\Tp|$ denotes the number of nodes in $\Tp$.
	
\end{lemma}

\begin{proof} If node $u\in \mathcal{T}'$ is not a leaf of $\mathcal{T}'$, we have $\deg_{\mathcal{T}'}(u)\ge 1$ and so
\[\deg_{\mathcal{T}}(u) - \deg_{\mathcal{T}'}(u)\le \deg_{\mathcal{T}}(u) - 1 \le \Delta-1\,.\]
If $u\in \mathcal{T}'$ is a leaf of $\mathcal{T}'$, we have
\[\deg_{\mathcal{T}}(u) - \deg_{\mathcal{T}'}(u) =  \deg_{\mathcal{T}}(u)  \le \Delta\,.\]

We therefore have
	\begin{align*}
	V(\mathcal{\Tp}) & = \sum_{u \in \Tp}\left( \deg_{\mathcal{T}'}(u) - \deg_{\mathcal{T'}}(u)\right) -1\\
	&\leq |\mathcal{L}_{\mathcal{T}'}| \cdot \Delta + (|\Tp| - |\mathcal{L}_{\mathcal{T}'}|)(\Delta - 1) - 1\\
	&=|\Tp|  \cdot (\Delta-1) + |\mathcal{L}_{\mathcal{T}'}| - 1\\
	& \leq \frac{(\Delta^2 - 1)\cdot |\Tp| + 1}{\Delta} - 1\\
	& \leq \left(\Delta - \frac{1}{\Delta}\right)|\Tp|
	, \label{eq:Part1} \tag{$1$}
	\end{align*}
	where the second inequality follows from Lemma~\ref{lemma:leafbound}.
\end{proof}	

\section{Data generating process for regression simulation}
\label{sec:simulation-regression}

We first form a balanced 3-regular tree with $p=243$ leaves. We express the tree by a binary matrix $\boldsymbol{A}\in\{0,1\}^{p\times |\mathcal{T}|}$ with rows corresponding to features and columns corresponding to nodes. Each entry $A_{ju}=1$ if node $u$ is an ancestor of leaf $j$ or if $u=j$, and $A_{ju}=0$ otherwise.  
For a given $K$, we cut the tree into $K$ subtrees. The roots of the subtrees form $\mathcal{B}^*$. We want to set the coefficients corresponding to the leaves within each subtree to the same value. To achieve this, we generate a vector of length $K$, denoted as $\widetilde{\btheta}^*$, with the first $(1 - \beta)K$ elements set to $0$; the other $\beta K$ elements of of $\widetilde{\btheta}^*$ are independently drawn from $\normal(0, 0.5^2)$.   Then we set $\btheta^* = \boldsymbol{A}_{\mathcal{B}^*} \widetilde{\btheta}^*$, where 
$\boldsymbol{A}_{\mathcal{B}^*}$ is matrix $\boldsymbol{A}$ restricted to columns that correspond to the nodes in $\mathcal{B}^*$. Note that the columns of $\boldsymbol{A}_{\mathcal{B}^*}$ have disjoint supports as no two nodes in $\mathcal{B}^*$ can share a same descendant. Parameter $\beta$ controls the sparsity of $\widetilde{\btheta}^*$, and therefore sparsity of ${\btheta^*}$.

To simulate a setting with rare feature, we consider a design matrix
$\boldsymbol{X} := \widetilde{\boldsymbol{X}}\odot \boldsymbol{W} \in
\mathbb{R}^{n \times p} $ from a Bernoulli-Gaussian distribution.  The
entries $\widetilde{X}_{ij}$ are generated i.i.d from standard normal
distribution.  The entries $W_{ij}$ are drawn i.i.d from $\text{\rm
  Bernoulli}(\rho)$. The Bernoulli parameter $\rho$ determines the
level of rareness in the design matrix. Also $\odot$ represents the
entry-wise product of two matrices. Finally, the high-dimensional
linear model is generated by
\begin{align}
\by=\bX \btheta^*+\beps,\quad \beps\sim \normal(\boldsymbol{0},
  \sigma^2\boldsymbol{I}_n),\label{eq:lm}
\end{align}
where $\sigma = c \frac{\left\Vert\boldsymbol{X} \btheta^* \right\Vert_2}{\sqrt{n}}$. We fix the parameters as $ n=100, p=243, \beta= 0.6, \rho = 0.2,  \sigma = 0.6$, and vary $K$ from 21 to 93.

\section{Stocks data}
\label{sec:companies}

Table \ref{table:aggregation_companies} shows the achieved
aggregation result.

\begin{landscape}
\begin{table}[]
	\caption{\em\small The table shows $40$ clusters that are aggregated by our procedure. The rightmost column is the mean log-volatility. The column Company/Number of Companies shows how many companies are in each cluster (or the name of the company when there is only one). The columns from Sector to Industry National show the classification of the clusters that are selected.}
	\label{table:aggregation_companies}
        \resizebox{1.4\textwidth}{!}{%
          \begin{tabular}{|l|l|l|l|l|l|l|}
\hline
Sector & Subsector & Industry Group & Industry & Industry National & Company/Number of Companies & Mean \\ \hline
\multicolumn{5}{|l|}{Agriculture, Forestry, Fishing and Hunting} & 5 & -7.69 \\ \hline
\multicolumn{5}{|l|}{Mining, Quarrying, and Oil and Gas Extraction} & 63 & -6.51 \\ \hline
\multicolumn{5}{|l|}{Utilities} & 28 & -8.4 \\ \hline
\multicolumn{5}{|l|}{Construction} & 45 & -7.55 \\ \hline
\multicolumn{5}{|l|}{Manufacturing I} & 73 & -8.03 \\ \hline
\multicolumn{5}{|l|}{Manufacturing III} & 576 & -7.48 \\ \hline
\multicolumn{5}{|l|}{Wholesale Trade} & 75 & -7.62 \\ \hline
\multicolumn{5}{|l|}{Retail Trade I} & 75 & -7.56 \\ \hline
\multicolumn{5}{|l|}{Retail Trade II} & 34 & -7.5 \\ \hline
\multicolumn{5}{|l|}{Transportation} & 44 & -7.81 \\ \hline
\multicolumn{5}{|l|}{Warehousing} & 3 & -8.31 \\ \hline
\multicolumn{5}{|l|}{Information} & 261 & -7.5 \\ \hline
\multicolumn{5}{|l|}{Real Estate and Rental and Leasing} & 50 & -7.34 \\ \hline
\multicolumn{5}{|l|}{Professional, Scientific, and Technical Services} & 77 & -7.55 \\ \hline
\multicolumn{5}{|l|}{Administrative and Support and Waste Management and Remediation Services} & 57 & -7.65 \\ \hline
\multicolumn{5}{|l|}{Educational Services} & 12 & -7.38 \\ \hline
\multicolumn{5}{|l|}{Health Care and Social Assistance} & 41 & -7.36 \\ \hline
\multicolumn{5}{|l|}{Arts, Entertainment, and Recreation} & 16 & -7.75 \\ \hline
\multicolumn{5}{|l|}{Accommodation and Food Services} & 52 & -7.86 \\ \hline
\multicolumn{5}{|l|}{Other Services (except Public Administration)} & 6 & -8.08 \\ \hline
\multicolumn{5}{|l|}{Nonclassifiable Establishments} & 9 & -7.28 \\ \hline
 & \multicolumn{4}{l|}{Wood Product Manufacturing} & 6 & -7.55 \\ \cline{2-7} 
 & \multicolumn{4}{l|}{Paper Manufacturing} & 19 & -8.14 \\ \cline{2-7} 
 & \multicolumn{4}{l|}{Printing and Related Support Activities} & 4 & -7.91 \\ \cline{2-7} 
 & \multicolumn{4}{l|}{Petroleum and Coal Products Manufacturing} & 16 & -7.93 \\ \cline{2-7} 
 & \multicolumn{4}{l|}{Plastics and Rubber Products Manufacturing} & 17 & -7.65 \\ \cline{2-7} 
 & \multicolumn{4}{l|}{Nonmetallic Mineral Product Manufacturing} & 11 & -7.44 \\ \cline{2-7} 
 &  & \multicolumn{3}{l|}{Basic Chemical Manufacturing} & 28 & -7.18 \\ \cline{3-7} 
 &  & \multicolumn{3}{l|}{Resin, Synthetic Rubber, and Artificial and Synthetic Fibers and Filaments Manufacturing} & 6 & -7.91 \\ \cline{3-7} 
 &  & \multicolumn{3}{l|}{Pesticide, Fertilizer, and Other Agricultural Chemical Manufacturing} & 6 & -7.14 \\ \cline{3-7} 
 &  & \multicolumn{3}{l|}{Pharmaceutical and Medicine Manufacturing} & 299 & -6.37 \\ \cline{3-7} 
 &  & \multicolumn{3}{l|}{Paint, Coating, and Adhesive Manufacturing} & 6 & -8.06 \\ \cline{3-7} 
 &  & \multicolumn{3}{l|}{Soap, Cleaning Compound, and Toilet Preparation Manufacturing} & 14 & -8.18 \\ \cline{3-7} 
\multirow{-13}{*}{Manufacturing II} & \multirow{-7}{*}{Chemical Manufacturing} & \multicolumn{3}{l|}{Other Chemical Product and Preparation Manufacturing} & 5 & -8.35 \\ \hline
 & \multicolumn{4}{l|}{Securities, Commodity Contracts, and Other Financial Investments and Related Activities} & 64 & -7.95 \\ \cline{2-7} 
 & \multicolumn{4}{l|}{Insurance Carriers and Related Activities} & 87 & -8.3 \\ \cline{2-7} 
 & \multicolumn{4}{l|}{Funds, Trusts, and Other Financial Vehicles} & MANHATTAN BRIDGE CAPITAL INC & -7.64 \\ \cline{2-7} 
 &  & \multicolumn{3}{l|}{Depository Credit Intermediation} & 306 & -8.27 \\ \cline{3-7} 
 &  & \multicolumn{3}{l|}{Nondepository Credit Intermediation} & 32 & -7.67 \\ \cline{3-7} 
\multirow{-6}{*}{Finance and Insurance} & \multirow{-3}{*}{Credit Intermediation and Related Activities} & \multicolumn{3}{l|}{Activities Related to Credit Intermediation} & 9 & -7.59 \\ \hline
\end{tabular}
}
\end{table}
\end{landscape}

\section{NYC taxi data}
\label{sec:taxis}

\begin{figure}
	\centering
	\includegraphics[width=0.7\textwidth]{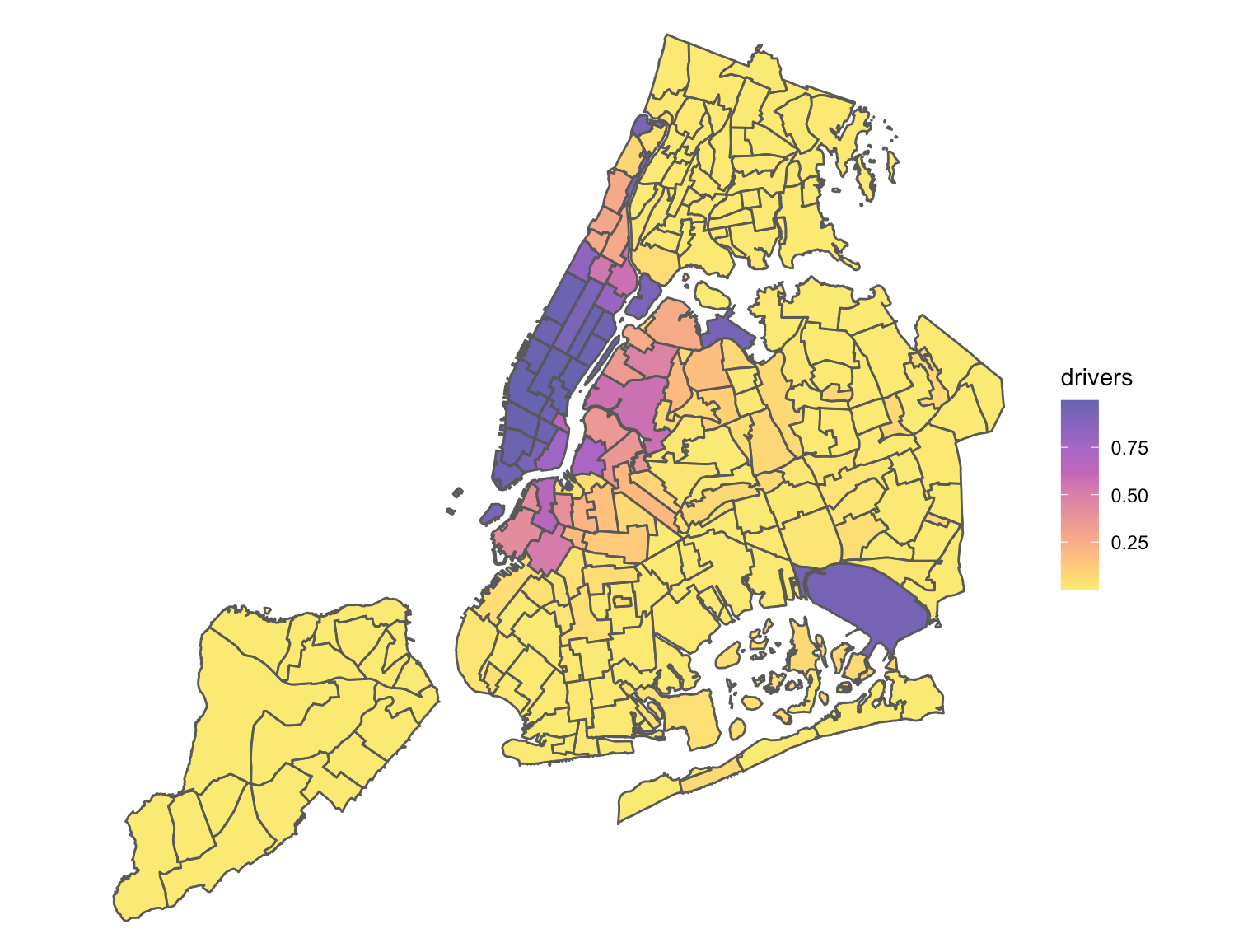}
	\caption{\em\small Map of neighborhoods colored with percentage of drivers who have started a trip from there. Most neighborhoods have fewer than 10\% of the drivers starting their trips there in the month of December 2013. }
	\label{fig:distribution_of_X}
\end{figure}

The
availability of taxis is not uniformly distributed across the city
(see Figure~\ref{fig:distribution_of_X}), and $\boldsymbol{X}$ is a highly sparse matrix: Most areas had
fewer than 10\% of the drivers starting their trips there during that
month, and in fact 109 out of the 194 neighborhoods have seen less
than 1\% of the drivers.

We study how the aggregation results vary with sample size. To do so, we randomly subset the original dataset to different sizes, and perform the above-mentioned procedure on each sample. The number of achieved groups for each sample size is shown in Table~\ref{table:changing_size}. As expected, reduced sample sizes leads to fewer rejections and therefore fewer aggregated groups.
\begin{table}
	\centering
	\caption{\em\small Achieved number of groups with decaying sample size.}
	\label{table:changing_size}
        \begin{tabular}{c|c|c}
Sample Size &  p & Number of Groups\\
\hline\hline
$n$ = 32704 & 194 & 44\\
$n/2$ = 16352 & 194 & 42\\
$n/4$ = 8176 & 194 & 29\\
$n/8$ = 4088 & 194 & 29\\
$n/16$ = 2044 & 194 & 18\\
$n/32$ = 1022 & 194 & 12\\
\hline
\end{tabular}

\end{table}

\subsection{FSR with synthetic data}
\label{sec:changing-size}

To directly evaluate the aggregation recovery performance of 
$\HAT$, we create a synthetic response based on the tree structure
$\mathcal{T}$ constructed by the neighborhoods, as well as the
observed trip counts data $\boldsymbol{X}$. In addition, we take the
aggregation result and fitted coefficients from
Section~\ref{sec:taxi-aggregation} as the true aggregation and true
vector $\btheta^*$. We simulate the response $100$ times independently
according to \eqref{eq:lm} with $\sigma = 15$.  We use the same debiased method to calculate the node-wise $p$-values and perform our testing procedure with target $\FSR$ levels varying from $\alpha = 0.01$ to $ \alpha =0.3$.  We compare the aggregation results with the true aggregation and compute $\FSR$ and average power over the $100$ runs. The results are shown in Table~\ref{table:synthetic}.

\begin{table}
	\centering
	\caption{\em\small Achieved FSR and average power by our algorithm with synthetic data where noise level is $\sigma = 15$.}
	\label{table:synthetic}
\begin{tabular}{c|c |c}
Target Level & FSR & Average Power\\
\hline\hline
0.01 & 0.000 & 0.547\\
0.02 & 0.000 & 0.560\\
0.05 & 0.000 & 0.577\\
0.10 & 0.001 & 0.593\\
0.20 & 0.003 & 0.608\\
0.30 & 0.003 & 0.620\\
0.40 & 0.004 & 0.626\\
0.50 & 0.005 & 0.632\\
\hline
\end{tabular}

\end{table}

\bibliography{references}

\end{document}